\documentclass[reprint,english,superscriptaddress,preprintnumbers,amsmath,amssymb,aps,prd,nofootinbib]{revtex4-2}

\usepackage[utf8]{inputenc}
\usepackage{diagbox}
\usepackage{orcidlink}
\usepackage{tikz,xcolor,hyperref}
\usepackage{mathtools}
\usepackage{amsfonts}
\usepackage{mathrsfs}
\usepackage{bbm}
\usepackage{slashed}
\usepackage{graphicx}
\usepackage{color}
\usepackage{array}

\usepackage{booktabs}
\usepackage{makecell}
\usepackage{epstopdf}
\usepackage[caption=false]{subfig}

\usepackage{xspace}
\usepackage{siunitx}
\usepackage{hyperref}
\usepackage[nameinlink]{cleveref}

\usepackage{xifthen}
\usepackage{xcolor}
\hypersetup{
	colorlinks,
	linkcolor={blue!75!black},
	citecolor={blue!75!black},
	urlcolor={blue!75!black}
}

\setkeys{Gin}{width=0.48\textwidth}

\newcommand{\bs}[1]{\boldsymbol{#1}}

\graphicspath{
	{./figures/}
	{../figures/}
}

\def\eq#1{(\ref{#1})}

\def\Eq#1{Eq.~(\ref{#1})}

\def\Fig#1{Fig.~\ref{#1}}

\def\Sec#1{Sec.~\ref{#1}}
\def\App#1{App.~\ref{#1}}

\newcommand{\gettitle}{The QCD moat regime and its real-time properties}

\hypersetup{
	colorlinks,
	linkcolor={red!75!black},
	citecolor={blue!75!black},
	urlcolor={blue!75!black}, 
	pdftitle={\gettitle},
	pdfauthor={},
	pdfkeywords={}
	{} 
	bookmarksopen=true,
	bookmarksopenlevel=2,
	bookmarksnumbered=true
}

\begin{document}
\title{\gettitle}

\author{Wei-jie Fu \,\orcidlink{0000-0002-5647-5246}}
\affiliation{School of Physics, Dalian University of Technology, Dalian, 116024, P.R. China}
  
\author{Jan M. Pawlowski\,\orcidlink{0000-0003-0003-7180}}
\affiliation{Institut f\"ur Theoretische Physik, Universit\"at Heidelberg, Philosophenweg 16, 69120 Heidelberg, Germany}
\affiliation{ExtreMe Matter Institute EMMI, GSI, Planckstra{\ss}e 1, D-64291 Darmstadt, Germany}

\author{Robert D.\ Pisarski \,\orcidlink{0000-0002-7862-4759}}
\affiliation{Department of Physics, Brookhaven National Laboratory, Upton, New York 11973, USA}

\author{Fabian Rennecke \,\orcidlink{0000-0003-1448-677X}}
\affiliation{Institut f\"ur Theoretische Physik, Justus-Liebig-Universit\"at Gie\ss en, 35392 Gie\ss en, Germany}
\affiliation{Helmholtz Research Academy Hesse for FAIR, Campus Gie\ss en, 35392 Gie\ss en, Germany}

\author{Rui Wen \,\orcidlink{0000-0002-1319-1331}}
\affiliation{School  of  Nuclear  Science  and  Technology, University  of  Chinese  Academy  of  Sciences,  Beijing,  P.R.China  100049}	

\author{Shi Yin \,\orcidlink{0000-0001-5279-6926}}
\email{Shi.Yin@theo.physik.uni-giessen.de}
\affiliation{Institut f\"ur Theoretische Physik, Justus-Liebig-Universit\"at Gie\ss en, 35392 Gie\ss en, Germany}

	
\begin{abstract}
Dense QCD matter may exhibit crystalline phases. Their existence is reflected in a moat regime, where mesonic correlations feature spatial modulations. We study the real-time properties of pions at finite temperature and density in QCD in order to elucidate the nature of this regime. We show that the moat regime arises from particle-hole-like fluctuations near the Fermi surface. This gives rise to a characteristic peak in the spectral function of the pion at nonzero spacelike momentum. This peak can be interpreted as a new quasi particle, the moaton. In addition, our framework also allows us to directly test the stability of the homogeneous chiral phase against the formation of an inhomogeneous condensate in QCD. We find that the formation of such a phase is highly unlikely for baryon chemical potentials $\mu_B \leq 630$\,MeV.    
\end{abstract}

\maketitle
	
\section{Introduction}

The phase structure of QCD at finite density is largely unknown. Results from lattice gauge theory show that there is a smooth chiral crossover for $\mu_B/T \lesssim 3$ \cite{HotQCD:2018pds,Borsanyi:2020fev}, where $\mu_B$ and $T$ are the baryon chemical potential and the temperature. Functional methods have predicted that this crossover ends in a critical endpoint (CEP) at around $(T,\mu_B) \approx (107,632)$~MeV \cite{Fu:2019hdw, Gao:2020fbl, Gunkel:2021oya}. These predictions from direct computations in QCD have subsequently been confirmed by various extrapolations of the available lattice data, e.g., \cite{Basar:2023nkp, Hippert:2023bel, Clarke:2024ugt, Shah:2024img}.

The CEP is extensively searched for in heavy-ion collision experiments \cite{Luo:2017faz, Bzdak:2019pkr}. However, to date there are no indications for its existence, as the available data do not show any significant deviations from the noncritical baseline for collision energies $\sqrt{s} \geq 7.7$~MeV, corresponding to $\mu_B \lesssim 400$~MeV \cite{STAR:BES2}. This is not surprising, as CEP signals typically rely on critical scaling, yet scaling can only be observed very close to the phase transition; see, e.g., \cite{Braun:2010vd, Schaefer:2011ex, Fu:2019hdw, Braun:2020ada, Gao:2021vsf, Fu:2023lcm, Bernhardt:2023hpr, Braun:2023qak, Lu:2023mkn}.

\begin{figure}[t]
\includegraphics[width=0.4\textwidth]{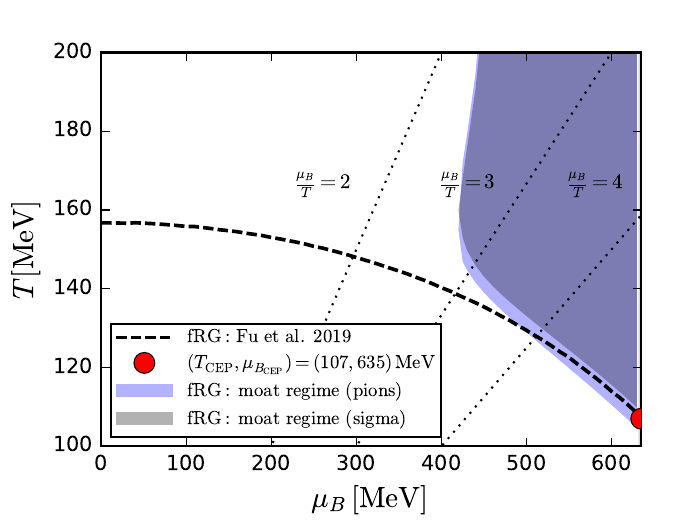}\hspace{0.5cm}
\caption{The phase boundary of the chiral phase transition of QCD \cite{Fu:2019hdw} together with the moat regime identified though pion and sigma correlations.}
\label{fig:phase_moat}
\end{figure}

In contrast, in Ref.\ \cite{Fu:2019hdw} a region with negative spatial wave functions of mesonic correlations has been found in QCD using the functional renormalization group (fRG). This implies the existence of a moat regime, where the static dispersion of mesons has a minimum at nonzero spatial momentum $p_M = |\boldsymbol{p}_M| >0$. $p_M$ corresponds to the wave number of an underlying spatial modulation. As shown in Fig.~\ref{fig:phase_moat}, the moat regime covers a large region of the phase diagram. While the moat regime itself has not received much attention until recently \cite{Pisarski:2021qof, Rennecke:2021ovl}, it turns out to be a generic feature of systems with spatially modulated phases, ranging, e.g., from condensed matter systems \cite{Yoshimori, FFLO1, Hornreich:1975, Braz, Elihu, Amnon, Braz2, Selke, qhstr1, fogler1, chalker, qhstr2, QHE, qhstr3, compete_GL, Chakrabarty:2011, modulation2, Sedrakyan:2014,  LC2, Selinger_2022, Pu:2024pwz}, various effective models that share some features with QCD \cite{Buballa:2014tba, Pisarski:2018bct, Pisarski:2020dnx, Koenigstein:2023yzv, Motta:2024agi, Pannullo:2024sov, Motta:2024rvk}, and, also including, systems with a symmetry under combined charge and complex conjugation \cite{vanBaal:2000zc, Nishimura:2014rxa, Reinosa:2014ooa, Schindler:2019ugo, Schindler:2021otf, Haensch:2023sig, Winstel:2024dqu}, to pattern formation in general \cite{cross1993pattern, Seul:1995}.

%
%

Owing to the large size of the moat regime, it is conceivable that the matter created in intermediate-energy heavy-ion collisions can spend an appreciable amount of its lifetime in this regime. In this case, it has been argued that particle production is affected \cite{Pisarski:2020gkx}. Indeed, correlations generated through thermodynamic fluctuations \cite{Pisarski:2021qof} and pion interference \cite{Rennecke:2023xhc} show characteristic peaks at $p_M$ for moaton quasiparticles, i.e.\ pions in a moat regime, which are the softest modes in the system for $|\boldsymbol{p}| \approx p_M$. In addition, dilepton production is enhanced around the moaton production threshold \cite{Nussinov:2024erh}. Similar signals have been found for specific inhomogeneous phases \cite{Fukushima:2023tpv, Hayashi:2024sae}.

Much like transport phenomena, the observables for a moat regime developed in Refs.~\cite{Pisarski:2021qof, Rennecke:2023xhc, Nussinov:2024erh} require knowledge of the real-time spectral properties of particles in a moat regime.
So far, explicit calculations relied on the assumption of moatons with a quasiparticle spectral function $\rho(p_0,\bs{p}^2) \sim \delta\big(p_0^2-E_M^2(\bs{p}^2)\big)$, where $E_M(\bs{p}^2)$ has a global minimum at $p_M$. This may, however, be an oversimplification of the realistic spectrum.

The phase structure, and more generally thermodynamic properties, are equilibrium concepts that are accessible at imaginary time. Consequently, they are related to the \emph{spacelike} properties of matter. For example, the formation of condensates is fully characterized by Euclidean correlation functions. In addition, second order phase transitiosn, e.g., at the CEP, are signaled by a diverging correlation length, which is given by the inverse spacelike \emph{screening} mass of the critical mode \cite{Zinn-Justin:2002ecy}. It is hence no surprise that the massless mode at the CEP appears in the spacelike region of the scalar meson
spectral function \cite{Fujii:2003bz, Yokota:2016tip}. To be more precise, the critical mode of the CEP is a mixture of the chiral condensate, the density and the Polyakov loops \cite{Haensch:2023sig}.

In general, the instability of the ground state of a system towards a different phase is always signaled by a sign change of the \emph{static} propagator \cite{Weinberg:1987vp}. A relevant example here is the possible instability of a homogeneous ground state against inhomogeneous condensation \cite{Buballa:2014tba}. This instability is generated by particle-hole fluctuations around the Fermi surface, see, e.g., \cite{Jeong:2024rst}, and occurs if the static moaton dispersion vanishes at the minimum, $E_M(p_M) = 0$ \cite{Rennecke:2021ovl}.

For broken Lorentz symmetry, there is no simple relation between the spacelike and timelike properties of matter. Hence, a simplistic quasiparticle picture as mentioned above might not be warranted. In particular since the existence of a moaton quasiparticle has not been established yet.
And while the phase structure is determined by the spacelike properties of particles, their real-time properties are necessary to understand transport phenomena and experimental signals. In order to fully understand, and potentially also measure, the moat regime, we hence need to know its real-time properties. To this end, in this paper we compute the spectral function of pions in the moat regime based on direct calculations in QCD with the fRG \cite{Fu:2019hdw}. 

A by-product of this analysis is that we can directly test the stability of the system against inhomogeneous fluctuations at finite density. In this work, we therefore also put forward a direct way to perform a stability analysis in QCD, which is complementary to the recently developed methods based on the 2PI effective action \cite{Motta:2024agi} and the chiral susceptibility \cite{Motta:2024rvk}.

\section{Spatial modulations and the meson spectral function}\label{sec:spam}

In the moat regime correlation function features spatial modulations. This does not necessarily mean that the system is in an inhomogeneous phase, where some of the spatial symmetries are broken spontaneously. Even in the absence of long-range crystalline order, correlations functions can still exhibit periodic oscillations. A well-known example are Friedel oscillations \cite{fetter2012quantum, Kapusta:1988fi,DiazAlonso:1989up,Diaz-Alonso:1998eva,Liu:2007bu,Mu:2008ic}, where periodic oscillations in bosonic two-point functions arise from fermionic particle-hole fluctuations around a sharp Fermi surface at $T=0$. However, as shown in Ref.\ \cite{SYFRMoat}, the moat regime is a qualitatively different phenomenon that entails spatial oscillations at both zero and nonzero temperatures. In QCD, it can be characterized through the momentum dependence of mesonic two-point functions \cite{Fu:2019hdw}.

The meson propagator is given by the connected two-point function 
\begin{align}
    G_\phi(p) = \langle \phi(p)\phi(-p)\rangle_c \,. 
\label{eq:MesonProp} 
\end{align}
It can be expressed in terms of the QCD effective action $\Gamma[\Phi]$ in the presence of emergent mesonic composites, where $\Phi=(q,\bar q,A,c,\bar c, \bs{\pi}, \sigma,\dots)$ and the propagator is given by 
\begin{align}
    G_\phi(p)(2 \pi)^4\delta(p+q) = \left(\frac{ \delta^2 \Gamma[\Phi]}{\delta\Phi\delta\Phi}\right)^{-1}_{\phi\phi,\Phi_\textrm{EoM}}(p,q)\,.
\label{eq:MesonProp} 
\end{align}
The propagator is evaluated on the solution $\Phi_\textrm{EoM}$ of the quantum equations of motion (EoM), $\delta\Gamma[\Phi]/\delta\Phi = 0$. This entails that the propagator is diagonal in momentum space. The subscript ${}_{\phi\phi}$ indicates the diagonal meson part of the propagator matrix. 

The standard QCD effective action is a functional of the fundamental fields, quarks, gluons and, in the present gauge-fixed setup, ghosts, $(q,\bar q,A,c,\bar c)$. Mesonic fields are defined through appropriate interpolating fields, e.g., for pions $\bs{\pi} \sim \bar q\, i \gamma_5 \bs{T}\, q$, where $\bs{T}$ are generators of an SU(2) flavor symmetry. They arise as dynamical degrees of freedom through resonances in four-quark interactions, and can be captured by the emergent composites approach (also called dynamical hadronization) \cite{Gies:2001nw, Gies:2002hq, Pawlowski:2005xe, Floerchinger:2009uf, Fu:2019hdw, Fukushima:2021ctq}. This allows us to treat emergent bound states together with fundamental fields in a consistent way, and the respective effective action in general depends on $\Phi=(q,\bar q,A,c,\bar c, \bs{\pi}, \sigma,\dots)$, where $\sigma$ is the scalar meson and the dots stands for further hadronic fields. On the equations of motion of the composite fields, the effective action reduces to the standard one of QCD, see Ref.\ \cite{Fu:2019hdw} and the next section for more details. 

At finite temperature and density, the full Euclidean meson propagator \labelcref{eq:MesonProp} can in general be parametrized as  
\begin{align}
  G_\phi(p) &=\frac{1}{Z_{\phi}^{\parallel}(p_0, \bs{p})\,\left( p_0^2+m^2_{\phi}\right) +Z_{\phi}^{\perp}(p_0, \bs{p})\,\bs{p}^2 }\,,
  \label{eq:Gam2phiphiB}
\end{align}
with $p=(p_0, \boldsymbol{p})$. We define the O(4) field $\phi=(\sigma, \bs{\pi})$ for convenience.
The temporal and spatial wave functions, $Z_{\phi}^{\parallel}$ and $Z_{\phi}^{\perp}$, are in general unequal due to broken Lorentz invariance in a medium. 
The meson mass $m_{\phi}$ is the physical pole mass defined by  $G^{-1}_\phi(p_0^2=-m_\phi^2,\bs{p}=0)=0$. We note in passing that the wave functions also incorporate the wave function renormalization (its value at the RG point), and hence are commonly identified with the latter.

A simple way to see how the moat regime manifests itself here is to consider the static propagator, $G_\phi(\bs{p}) = G_\phi(p)|_{p_0=0}$. A spatial modulation of the two-point correlation $G_\phi(\bs{x})$ with wave number $\bs{p}_M$ is equivalent to $G_\phi(\bs{p})$ being peaked at $\bs{p}_M$. If we assume, just for illustration, that the wave functions depend only mildly on the spatial momentum, $Z_\phi^{\parallel/\perp}(\bs{p}^2) \approx Z_0^{\parallel/\perp} + Z_1^{\parallel/\perp}\bs{p}^2$, and that the system retains rotational invariance, spatial modulations require $(G_\phi^{-1})^\prime(\bs{p}_M^2) = 0$ with 
\begin{align}\label{eq:pmoat}
    \bs{p}_M^2 = -\frac{Z_0^\perp}{2Z_1^\perp} - \frac{Z_1^\parallel m_\phi^2}{2Z_1^\perp}\,.
\end{align}
Causality implies $Z_1^\parallel \geq 0$ and stability of the solution of the EoM requires $Z_1^\perp > 0$, so the static propagator is finite and positive for all $\bs{p}$ \cite{Weinberg:1987vp}. Thus, to have a nonzero wave number $\bs{p}_M^2 > 0$, $Z_0^\perp = Z_\phi^\perp(0,\bs{0})$ needs to become sufficiently negative. This has been found in \cite{Fu:2019hdw, Koenigstein:2021llr, Pannullo:2024sov} and is the basis for effective models for moaton quasiparticles, Refs.\ \cite{Pisarski:2020dnx, Pisarski:2021qof, Rennecke:2021ovl, Rennecke:2023xhc, Nussinov:2024erh}.

In this approximation the propagator can be expressed as
\begin{align}
  G_\phi(p) \approx\frac{1}{\big(Z_0^{\parallel} + Z_1^{\parallel} \bs{p}^2 \big)\, p_0^2 + Z_1^\perp \big(\bs{p}^2 - \bs{p}_M^2\big)^2 + m_{\rm eff}^2}\,,
  \label{eq:Gam2phiphiB}
\end{align}
with
\begin{align}\label{eq:meff}
    m_{\rm eff}^2 = Z_0^{\parallel} m_\phi^2 - Z_1^\perp \bs{p}_M^4\,.
\end{align}
From this we can read-off the static meson dispersion
\begin{align}\label{eq:Estat}
    E_\phi^{\rm(stat)}(\bs{p}) \equiv \sqrt{G_\phi^{-1}(\bs{p})} = \sqrt{Z_1^\perp \big(\bs{p}^2 - \bs{p}_M^2\big)^2 + m_{\rm eff}^2}\,. 
\end{align}
At $\bs{p} = \bs{p}_M$ the static dispersion is minimal,
\begin{align}\label{eq:Estat}
   E_\phi^{\rm(stat)}(\bs{p}_M) = m_{\rm eff} \leq m_\phi\,.
\end{align}
This shows that the energy gap of static mesons is lowered in the moat regime. This is the basis for the prediction of its experimental signatures \cite{Pisarski:2020gkx, Pisarski:2021qof, Rennecke:2023xhc, Nussinov:2024erh}.

\begin{figure}[t]
\includegraphics[width=0.45\textwidth]{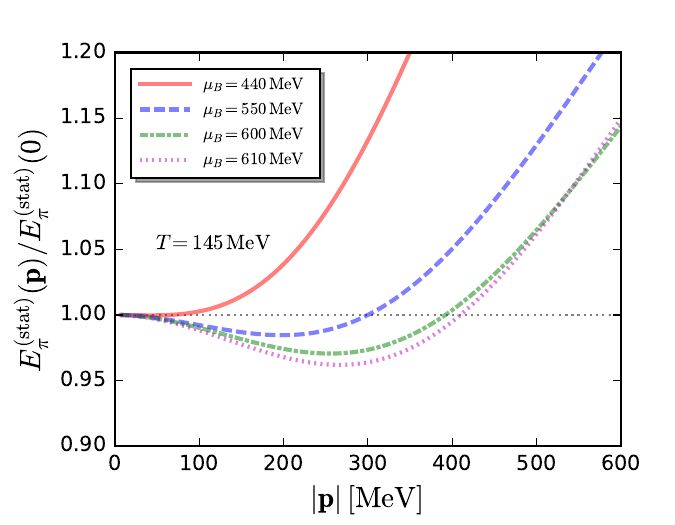}
\caption{Normalized static dispersion relation, defined in \Eq{eq:Estat}, for pions at $T=145$\,MeV and different $\mu_B$.}\label{fig:disper-relation}
\end{figure}

To illustrate this, we show the static pion dispersion found in this work at a fixed $T$ for different $\mu_B$ in \Fig{fig:disper-relation}. Since the meson mass $m_\phi$ depends on $T$ and $\mu_B$ in our full computations, we normalized the energy by its value at zero momentum for clarity. One clearly sees the moat regime at large $\mu_B$ through the nonmonotonic $|\bs{p}|$-dependence of the dispersion. We find that $|\bs{p}_M|$ quickly obtains values on the order of a few hundred MeV. We will quantify this in more detail below.   

Furthermore, from \Eq{eq:meff} follows that if 
\begin{align}\label{eq:inst}
    Z_1^\perp \bs{p}_M^4 \geq m_\phi^2\,,
\end{align}
this energy gap vanishes or even turns imaginary, which indicates that there is an instability towards the formation of a  different ground state \cite{Weinberg:1987vp}. Since this instability occurs at a nonzero spatial momentum $\bs{p}_M$, this favored ground state is likely an inhomogeneous phase \cite{Buballa:2014tba}. We will return to the discussion of inhomogeneous instabilities in QCD in \Sec{sec:stabilityana}. 

%
%

This example illustrates how spatial modulations, the moat regime and also inhomogeneous instabilities manifest themselves in the static propagator. However, to fully understand the moat regime also beyond the static limit, and in particular at real time, we need to know how the full frequency and momentum dependence of the real-time correlation function in the moat regime.

To this end, one may directly compute in real time, or, if the Euclidean propagator is already available, one may perform an analytic continuation from imaginary to real time via $p_0\to -\mathrm{i}(\omega\pm\mathrm{i}\epsilon)$; for applications with the fRG see, e.g., 
~\cite{Gasenzer:2007za, Gasenzer:2010rq, Floerchinger:2011sc, Strodthoff:2011tz, Kamikado:2013sia, Tripolt:2013jra, Pawlowski:2015mia, Yokota:2016tip, Kamikado:2016chk, Jung:2016yxl, Pawlowski:2017gxj, Yokota:2017uzu, Wang:2017vis, Tripolt:2018jre, Tripolt:2018qvi, Corell:2019jxh, Huelsmann:2020xcy, Jung:2021ipc, Tan:2021zid, Heller:2021wan, Fehre:2021eob, Roth:2021nrd, Roth:2023wbp, Horak:2023hkp} as well as the comprehensive review \cite{Dupuis:2020fhh} and references therein. In the latter case, the retarded meson propagator is obtained from \Eq{eq:Gam2phiphiB} by
\begin{align}
  G^R_\phi(\omega,\bs{p})=&\lim_{\epsilon\to 0^+} G_\phi\big(-\mathrm{i}(\omega+\mathrm{i}\epsilon),\bs{p}\big)\,.
  \label{eq:AnalyContiGamphi2}
\end{align}
From the retarded propagator we can directly extract the spectral function,
\begin{align}\label{eq:spec}
    \rho_\phi(\omega,\bs{p}) = \frac{1}{\pi} {\rm Im}\, G_\phi^R(\omega,\bs{p})\,.
\end{align}    
As explained above, in addition to its fundamental relevance, the spectral function is the central object for phenomenological studies of the moat regime and hence in the focus of this work. 

Following the arguments above, it seems natural to expect that the moat regime modifies the $\bs{p}$-dependence of the spectral function in the spacelike region at $\omega = 0$. This will be confirmed in the following, where we also investigate the behavior at $\omega > 0$.

\subsection{QCD spectral functions from the functional renormalisation group}\label{sec:FRG}

If the full effective action is known, we can follow the steps outlined above to obtain the spectral function. Here we compute the effective action from the generalised functional flow equation \cite{Pawlowski:2005xe, Fu:2019hdw}, 
\begin{align}\nonumber 
    \partial_t \Gamma_k[\Phi] &= \frac{1}{2}{\rm Tr} \big( G_k[\Phi]\, \left( \partial_t +\frac{\delta}{\delta \Phi} \dot\phi_k[\Phi] \right)  R_k\big)\\[2ex]
   &\hspace{1cm}- \int \dot\phi_{k,i}[\Phi] \bigg(\frac{\delta\Gamma_k[\Phi]}{\delta\phi_i}  + c_\sigma \delta_{\sigma i}\bigg)\,.
   \label{eq:flow}
\end{align}
The mean field is defined as $\Phi = \langle \hat\Phi\rangle$, where $\hat\Phi$ is the fluctuating quantum field in the path integral.
The terms proportional to $\dot \phi_k[\Phi]$ and its derivatives generate the emergent scalar-pseudoscalar mesons and are fixed by the requirement of absorbing the respective momentum channel in the four-quark scattering vertex, see the discussion around \Eq{eq:Gk}. This process in the fRG approach with emergent composites is also called dynamical hadronization. Note that, seemingly, the mean fields of the composites $\phi$ are scale-dependent, but they are independent fields in the effective action, which is obtained as a Legendre transformation also with respects to the currents of the composites. However, $\dot\phi[\Phi]$ reflects the scale dependence the underlying microscopic transformation of the basis in field space \cite{Pawlowski:2005xe}, 
\begin{align}
    \dot\phi_k[\Phi] = \langle \partial_t \hat\phi\rangle[\Phi]\,, 
\end{align}
for more details see \cite{Pawlowski:2005xe, Fu:2019hdw, Ihssen:2024ihp}. 

The regulator $R_k$ is a matrix in field space and suppresses the momentum modes of all fields with momenta $q^2 \!\lesssim\! k^2$, $t = \ln(k/\Lambda)$ is the RG time and $\Lambda$ is some reference scale. In the present work we use the initial RG scale as the reference scale and initialize the flow deep in the perturbative regime with $\Lambda = 20\, {\rm GeV}$. There we only have to fix the fundamental free parameters of QCD, i.e.\ the value of the strong coupling and the current quark masses. 
The full, field-dependent propagator in the presence of the regulator is similar to that at $k=0$ defined in \Eq{eq:MesonProp},
\begin{align}
    G_k[\Phi](p,q) = \left( \frac{\delta^2\Gamma_k[\Phi]}{\delta \Phi\delta\Phi} +R_k\right)^{-1}(p,q)\,,
\label{eq:Gk}
\end{align}
where we have suppressed all internal indices for the sake of simplicity. 

The scattering channel relevant for the formation of the $\sigma$ mode and the pions is the scalar-pseudoscalar four-quark interaction,
\begin{align}
    \lambda_{q,k} \bigg[ \frac{1}{4}(\bar q q)^2 +   (\bar q\, i \gamma_5 \bs{T}\, q)^2\bigg]\,,
\label{eq:4qc}
\end{align}
where $\lambda_{q,k}$ is an RG scale dependent (running) coupling.
By using that the RG flows of the composite fields are of the form $\langle \partial_t \hat\sigma_k \rangle \sim \bar q q$ and $\langle\partial_t  \hat{\bs{\pi}}_k \rangle \sim \bar q\, i \gamma_5\bs{T}\, q$, we completely encode this interaction for zero momentum exchange between the quark-antiquark pairs in the dynamics of pions and $\sigma$. For more details, including the approximations we use and a complete list of the flow equations that are solved here, we refer to Ref.\ \cite{Fu:2019hdw}.
%
\begin{figure}[t]
\includegraphics[width=1\columnwidth]{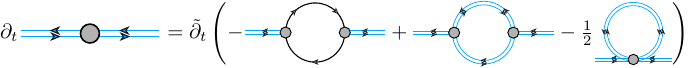}
\caption{Diagrammatic representation of the flow of the two-point function of mesons. Solid black lines are quarks and blue double-lines are mesons. The gray blobs denote the one-particle-irreducible (1PI) vertices. The derivative $\tilde \partial_t$ only hits the $k$-dependence of the regulator contribution to the propagators in \Eq{eq:Gk}. We emphasize that this equation is coupled to the corresponding flows of the quark propagator and the vertices as part of the fully coupled, nonperturbative QCD system \cite{Fu:2019hdw}.
}\label{fig:flowsGam2phiphi}
\end{figure}
%

The most relevant object for us here, the full meson propagator in \Eq{eq:Gam2phiphiB}, is obtained from the scale-dependent propagator at $k=0$ and evaluated on the solution of the EoM,
\begin{align}
   G_\phi(p) = G_{k=0}[\phi]\big|_{\rm EoM}(p)\,.
\end{align}
The diagrammatic representation of this flow is shown in \Fig{fig:flowsGam2phiphi}.
It is possible to use the analytic continuation described in the previous section to obtain the spectral function from this result. This is, however, an ill-conditioned inverse problem and hence necessitates some sort of reconstruction; see e.g.~\cite{Horak:2021syv, Pawlowski:2022zhh, Horak:2023xfb, Ali:2024xae} (Gau\ss ian Process regression), \cite{Jarrell:1996rrw, Asakawa:2000tr, Haas:2013hpa} (maximal entropy method), \cite{Burnier:2013nla, Rothkopf:2016luz}, (Bayesian inference), \cite{Ulybyshev:2017ped, Dudal:2019gvn, Dudal:2021gif} (Tikhonov regularisation), \cite{Fournier_2020, Yoon_2018, Kades:2019wtd, Zhou:2021bvw, Lechien:2022ieg} (neural networks), \cite{arsenault2016projected, Offler:2019eij} (kernel ridge regression), \cite{Cuniberti:2001hm, Burnier:2011jq, Cyrol:2018xeq, Fei_2021, fei2021analytical} (and basis expansions) for more discussions and the evaluation of different techniques. The staggering amount of different ones reflects indeed the ill-conditioned nature of the problem. 

The advantage of using the fRG is that we can analytically continue directly on the level of the flow $\partial_t G_k[\phi](p)$ \cite{Floerchinger:2011sc, Strodthoff:2011tz}. This way, we can compute the spectral function at $k=0$ without the need for any reconstruction. In practice, we do this in a two-step procedure. We solve the fully coupled QCD system without taking into account the full momentum dependence of the propagators, using in particular $Z_\phi^\perp = Z_0^\perp$ in the coupled system, cf.\ Ref.\ \cite{Fu:2019hdw}. We then use these results as input to separately solve the flow equation of the fully momentum dependent two-point functions shown in \Fig{fig:flowsGam2phiphi}, with the details given in \App{app:flow}. 
This procedure can be seen as the first step of an iteration towards the self-consistent inclusion of fully momentum dependent propagators, as done, e.g., in Ref.\ \cite{Helmboldt:2014iya}. We leave a fully self-consistent computation of the spectral functions in QCD to future work, but note that at least for O($N$) models this only leads to quantitative corrections \cite{Strodthoff:2016pxx, Horak:2023hkp, Jung:2021ipc}.

In any case, the full momentum dependence of the flow of the two-point function allows us to evaluate it also at complex frequencies, and we can directly perform the Wick rotation in \Eq{eq:AnalyContiGamphi2} to obtain the retarded propagator and, through \Eq{eq:spec}, the spectral function.


%
\section{The QCD moat regime}

In this section we will elaborate on the physics underlying the moat regime. First, we will show that it arises from particle-hole fluctuations. This can already be seen in Euclidean space. Second, we will present the first results on real-time correlations in the moat regime.  

\subsection{Particle-hole fluctuations and the moat regime}\label{sec:PH}

%
\begin{figure}[t]
\includegraphics[width=0.9\columnwidth]{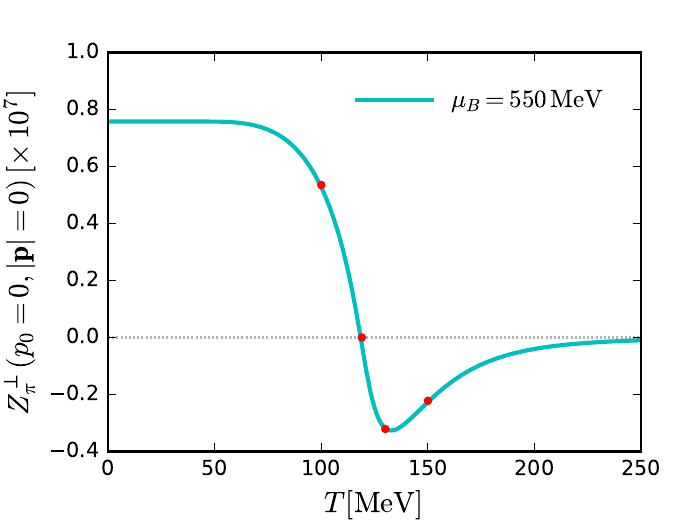}
\caption{Spatial pion wave functions $Z_{\pi}^{\perp}(p_0=0, |\boldsymbol{p}|=0)$ as a function of temperature $T$, at baryon chemical potential $\mu_B=550$ MeV. Red dots correspond to the temperatures in \Fig{fig:rhopi-3D-mub550}.}\label{fig:Zpion-T-mub550}
\end{figure}
%

In order to pin down the origin of the moat regime, we start by discussing our results in Euclidean space, where we solve the fully-coupled QCD system of Ref.\ \cite{Fu:2019hdw}.
As mentioned above, in this setup we use $Z_\phi^\perp = Z_0^\perp$, so that according to \Eq{eq:pmoat}, the moat regime is signaled directly by $Z_\phi^\perp < 0$ in this case. We focus on pions first, since they are, as the lightest degrees of freedom in the low-energy spectrum of QCD, physically most relevant.

As already reported in Ref.\ \cite{Fu:2019hdw}, we find that $Z_\pi^\perp$ becomes negative in a large region of the phase diagram. This is shown in
\Fig{fig:phase_moat}. We note that we restrict our analysis to $\mu_B \lesssim 630\,{\rm MeV}$ and $T \gtrsim 100\,{\rm MeV}$ because we only include the scalar-pseudoscalar quark scattering channel, \Eq{eq:4qc}, into our analysis. Other channels will become relevant beyond this regime \cite{Braun:2019aow}, and we want to focus on regions of the phase diagram where our systematic errors are under control. 

To illustrate the behavior of the spatial wave function within and outside the moat regime, we show $Z_\pi^\perp$ as a function of $T$ for $\mu_B = 550\,{\rm MeV}$ in \Fig{fig:Zpion-T-mub550}. For $T\gtrsim 120\, {\rm MeV}$, $Z_\pi^\perp$ is negative and approaches zero at large $T$. This approach to zero signals that mesons are turning into non-dynamical auxiliary fields and decouple from the QCD spectrum at large $T$ \cite{Braun:2014ata, Rennecke:2015eba}. We emphasize that this decoupling only happens in conjunction with $Z_\phi^\parallel$ also approaching zero. This is only true in the large-$T$ limit. At the zero crossing into the moat regime, $Z_\phi^\parallel \gg |Z_\phi^\perp|$, so quark correlations in the pion channel may be viewed as dynamical pions with a modified static dispersion. This will be clarified in the next section.

\begin{figure}[t]
\includegraphics[width=0.8\columnwidth]{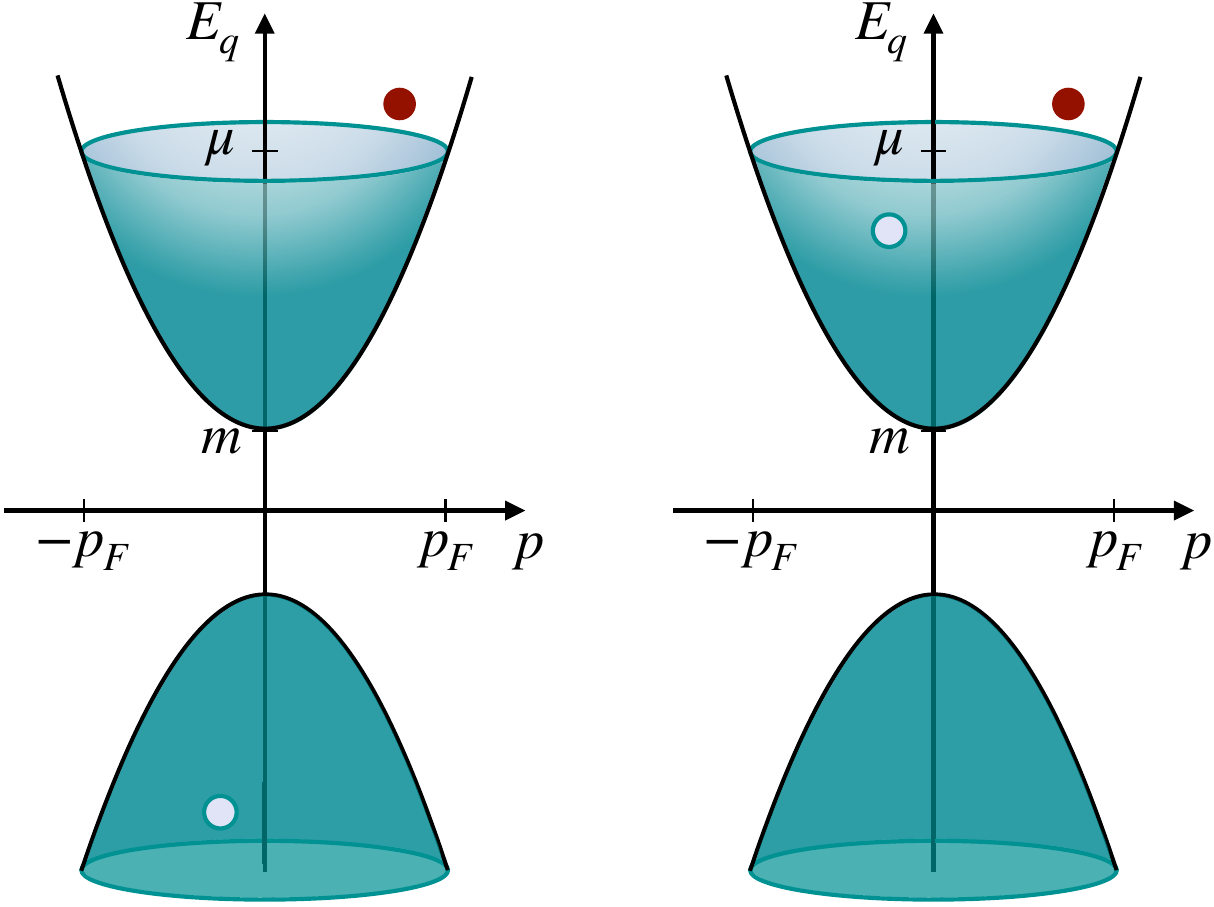}
\caption{Dirac cones illustrating the different quark processes contributing to the meson propagator. The Dirac cones reflect the quark dispersion $E_q(p)$. The green surface denotes the Dirac sea, which, at zero temperature, is filled up to the Fermi surface defined by $E_q(p_F) = \mu$, where $p_F$ is the Fermi momentum. At finite $T$, the Fermi surface is washed out, indicated by the fading color. The red dots denote quarks and the light dots either antiquarks (left) or quark-holes (right).\\
\emph{Left:} Creation-annihilation (CA) process involving fluctuations of a quark-antiquark pair. This process also includes the vacuum fluctuations of quarks.\\
\emph{Right:} Particle-hole (PH) process involving a quark--quark-hole pair. In the non-relativistic limit, the negative energy cone vanishes and only PH processes can occur.}\label{fs}
\end{figure}

We can gain further insights into the moat regime by investigating the different contributions to the pion propagator. Since we find the moat regime above the pseudocritical temperature of the chiral phase transition, the dominating contribution stems from quark fluctuations, cf.\ \Fig{fig:flowsGam2phiphi}. As discussed in more detail in \App{app:LandauDamping}, we can distinguish two manifestly different physical processes behind this contribution: fluctuations related to creation-annihilation (CA) processes of quarks, and particle-hole (PH) fluctuations.
They are illustrated in \Fig{fs}.

%
%

\begin{figure}[t]
\includegraphics[width=0.9\columnwidth]{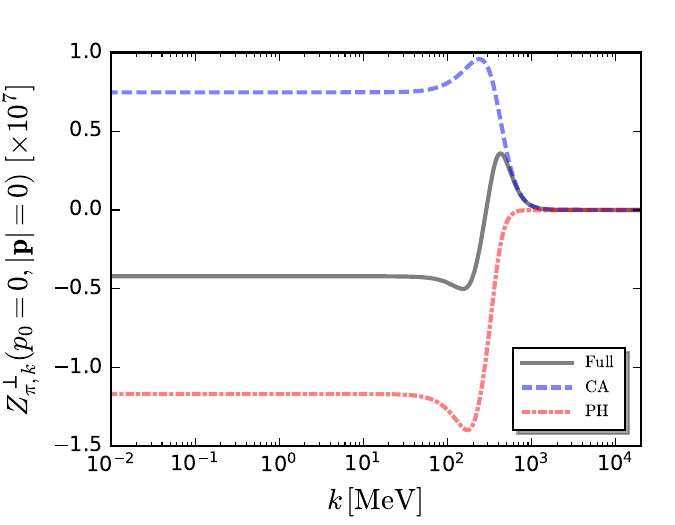}
\caption{Spatial pion wave function $Z_{\pi,k}^{\perp}(p_0=0, |\boldsymbol{p}|=0)$ as a function of the RG scale $k$ calculated at $T=120$ MeV and $\mu_B=600$ MeV. The solid black line shows the full result. The dashed blue line shows the contribution of creation-annihilation (CA) processes of quarks. The dot-dashed red line is the contribution of particle-hole (PH) fluctuations.}\label{fig:Zpion-k-LD}
\end{figure}

As demonstrated explicitly in \App{app:LandauDamping}, we can split the contribution of quarks to the RG flow of $Z_\pi^\perp$ into CA and PH processes. In \Fig{fig:Zpion-k-LD} we show the running of $Z_\pi^\perp$ in the moat regime at $\mu_B=600\,{\rm MeV}$ and $T=120\,{\rm MeV}$ together with the separate contributions from CA and PH processes. CA processes, which include the vacuum fluctuations of quarks, always give a positive contributions to $Z_\pi^\perp$. PH processes, on the other hand, can be negative and, as shown in \Fig{fig:Zpion-k-LD}, dominate over the CA processes in the moat regime. This shows that the moat regime is generated by particle-hole fluctuations of quarks in the medium.
Examples for the resulting static pion dispersion are shown in \Fig{fig:disper-relation}.

Furthermore, it follows from the discussion in \Sec{sec:spam} that for sufficiently negative $Z_\pi^\perp$ the system can become unstable towards the formation of an inhomogeneous condensate. Thus, the formation of inhomogeneous phases is also triggered by PH fluctuations. This may not be much of a surprise, as nonrelativistic condensed matter system can feature numerous crystalline phases, and CA process are absent in the nonrelativistic limit, cf.\ \Fig{fs}. Since we have shown in \Sec{sec:spam} that the moat regime is intimately related to spatially modulated phases, it is reassuring that the underlying mechanism of the formation of spatial modulations can work both in relativistic and nonrelativistic systems.

%
%

\subsection{Real-time correlations in the moat regime}\label{sec:moatspec}

%
\begin{figure*}[t]
\includegraphics[width=0.4\textwidth]{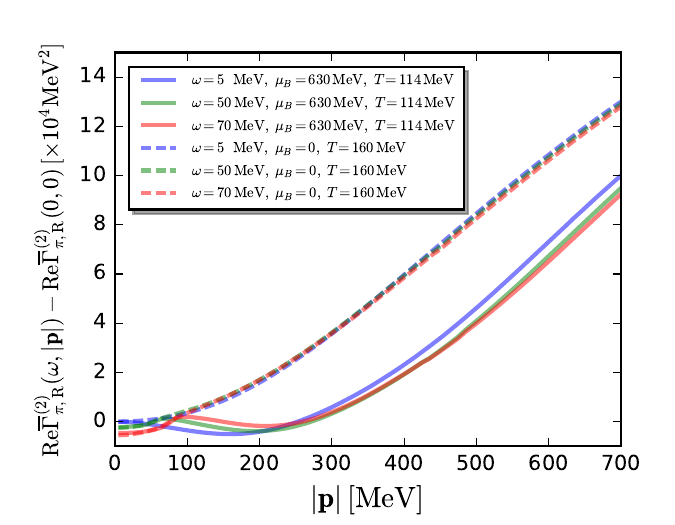}\hspace{0.5cm}
\includegraphics[width=0.4\textwidth]{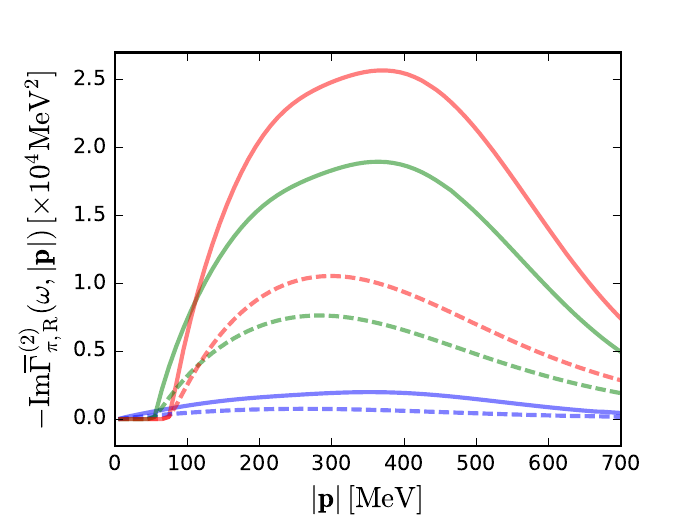}
\caption{Real (left panel) and imaginary (right panel) parts of the retarded two-point correlation function of pions as functions of spatial momentum for different frequencies. Results at large $\mu_B$ (solid lines) are in comparison to those at vanishing $\mu_B$ (dashed lines). Here the correlation function is normalized with $\bar{\Gamma}^{(2)}_{\pi, \mathrm{R}}=\Gamma^{(2)}_{\pi, \mathrm{R}}/Z^\perp_\pi(\omega=|\bs{p}|=0)|_{T=0,\mu_B=0}$.}\label{fig:Gam2R}
\end{figure*}
%

%
\begin{figure*}[t]
\includegraphics[width=0.4\textwidth]{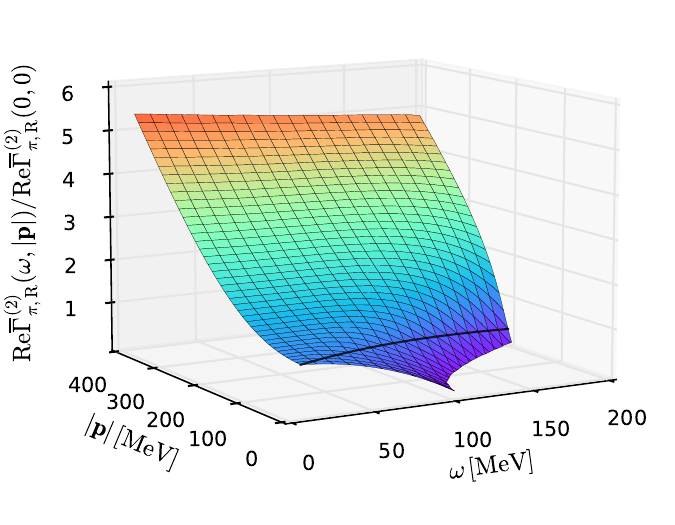}\hspace{0.5cm}
\includegraphics[width=0.4\textwidth]{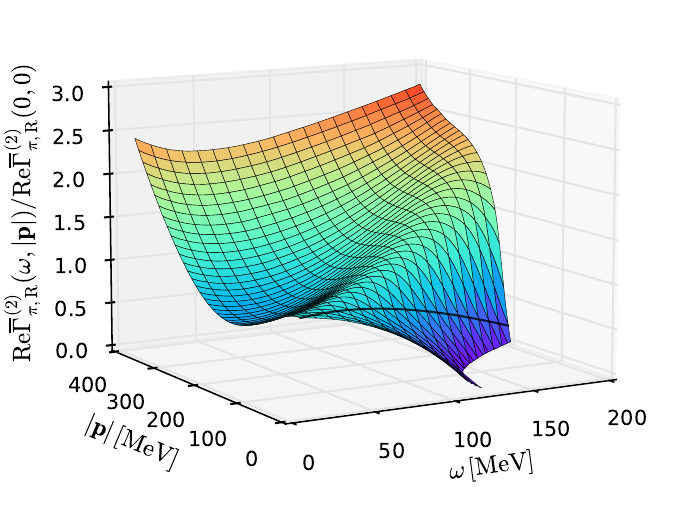}
\caption{3D plots of the real part of the retarded two-point correlation function of pions as a function of frequency and spatial momentum at $\mu_B=0$, $T=160$ MeV (left panel) and $\mu_B=630$, $T=114$ MeV (right panel). The black line indicates the light cone.}\label{fig:Gam2R-3D}
\end{figure*}
%

Now that we identified PH fluctuations as the origin of the moat regime, we study its real-time properties. To this end, we follow the steps described in \Sec{sec:spam} to obtain the retarded propagator and the spectral function of pions, Eqs.~\eq{eq:AnalyContiGamphi2} and \eq{eq:spec}, in QCD from analytically continued fRG flows.

In \Fig{fig:Gam2R} we show the real and imaginary parts of the retarded pion two-point function, $\Gamma_{\pi,R}^{(2)} = (G_\pi^R)^{-1}$, as functions of spatial momentum $|\bs{p}|$ for different real frequencies $\omega$ below the pion pole mass $m_\pi^{\rm pole}$. The two-point functions here are normalized by the vacuum pion wave function, i.e., $\bar{\Gamma}^{(2)}_{\pi, \mathrm{R}}=\Gamma^{(2)}_{\pi, \mathrm{R}}/Z^\perp_\pi(\omega=|\bs{p}|=0)|_{T=0,\mu_B=0}$.
This generalizes the results shown in \Fig{fig:disper-relation} to nonzero frequency.
The dashed lines are in the chirally restored phase outside the moat regime, and the solid lines are in the moat regime. We find a clear distinction between the timelike, $|\bs{p}| < \omega$, and the spacelike, $|\bs{p}| > \omega$, regions. As explained in \App{app:LandauDamping}, CA processes are only possible for timelike mesons, while PH processes occur in the spacelike region. This distinction hence highlights the differences in the contributions from CA and PH fluctuations. In the timelike region, ${\rm Re}\,\Gamma_{\pi,R}^{(2)}(\bs{p})$ is always increasing. Starting from the light cone $|\bs{p}| = \omega$, the behavior then changes qualitatively 
and we find substantial differences between the normal and the moat regime. In the normal regime, the real part is monotonically increasing with $|\bs{p}|$.

In contrast, one clearly sees in the left plot of \Fig{fig:Gam2R} the nonmonotonic $|\bs{p}|$-dependence of the static moat dispersion in the spacelike region.
This shows that the expected behavior for the static meson dispersion \eq{eq:Estat} in the normal and the moat regime persists also at nonzero $\omega$, at least for frequencies below $m_\pi^{\rm pole}$. We conclude that that the moat behavior induced by PH fluctuations discussed in the previous section at $\omega = 0$ extends to finite $\omega$ in the spacelike region of the pion two-point function.

The imaginary part ${\rm Im}\,\Gamma_{\pi,R}^{(2)}(\bs{p})$ shown in the right plot of \Fig{fig:Gam2R} is zero in the timelike region for $\omega < m_\pi^{\rm pole}$ both in the normal and the moat regime above the respective pseudocritical temperature. This is not surprising, as no ordinary scattering threshold is expected to occur at these low energies for pions. 
In contrast, as explained in \App{app:LandauDamping}, in the spacelike region an off-shell pion can create/annihilate a particle-hole pair of quarks or antiquarks. Since the quark and quark-hole are on-shell, this is kinematically allowed only for spacelike pions with $|\bs{p}| \geq w$. This is  \emph{Landau damping}, as it requires a quark/quark-hole state from the medium. This leads to a nonzero imaginary part in the spacelike region. While these contributions are qualitatively similar in the normal and a moat regime, we find a substantial enhancement of more than a factor of two in the magnitude of this contribution in the moat regime over the normal regime. We note that we compare the two-point function at different $T$ and $\mu_B$, so quantitative differences could also be attributed to these different conditions. However, as we will show below, the moat regime plays a significant role here. 

Since the moat regime manifests itself most clearly in the real part of the retarded two-point function, we show it as a function of both $\bs{p}$ and $\omega$ in \Fig{fig:Gam2R-3D} in the normal regime at $(T,\mu_B) = (160,0)$\,MeV (left) and the moat regime at $(T,\mu_B) = (114,630)$\,MeV (right). In the normal regime ${\rm Re}\,\Gamma_{\pi,R}^{(2)}(\bs{p})$ is a monotonically increasing function for all frequencies. In contrast, in the moat regime it is a nonmonotonic function for small and  intermediate frequencies. At very large frequencies, which are not shown here, ${\rm Re}\,\Gamma_{\pi,R}^{(2)}(\bs{p})$ is the same in both cases. This is to be expected, on the one hand, because the moat regime appears to be an in-medium effect and if the frequency scale is larger than the in-medium scale, all in-medium effects are negligible. On the other hand, the system is asymptotically free, so at very large $\omega$ the pion two-point function just looks like a weak quark correlation in the pion channel.  

We see from \Fig{fig:Gam2R-3D} that in the timelike region, there is no qualitative difference between the normal and the moat regime. At fixed $\omega$, ${\rm Re}\,\Gamma_{\pi,R}^{(2)}(\bs{p})$ is a monotonically increasing function in both cases. Crucially, the zero crossing of the real part of the two point function defines the ordinary, timelike dispersion relation of pions $E_\pi(\bs{p})$,
\begin{align}
    {\rm Re}\,\Gamma_{\pi,R}^{(2)}(\omega= E_\pi,\bs{p}) = 0\,,
\end{align}
and the pion pole mass is given by
\begin{align}
    m_\pi^{\rm pole} = E_\pi(\bs{0})\,.
\end{align}
\Fig{fig:Gam2R-3D} shows that $E_\pi(\bs{p})$ is a monotonically increasing function of $|\bs{p}|$ in both cases. We conclude that the pion in the timelike region just behaves like a normal pion, even in the moat regime.

%
\begin{figure*}[t]
\includegraphics[width=0.45\textwidth]{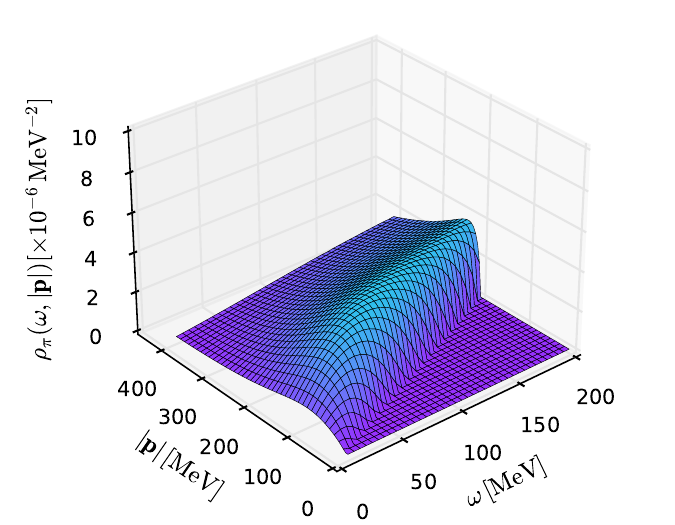}\hspace{0.5cm}
\includegraphics[width=0.45\textwidth]{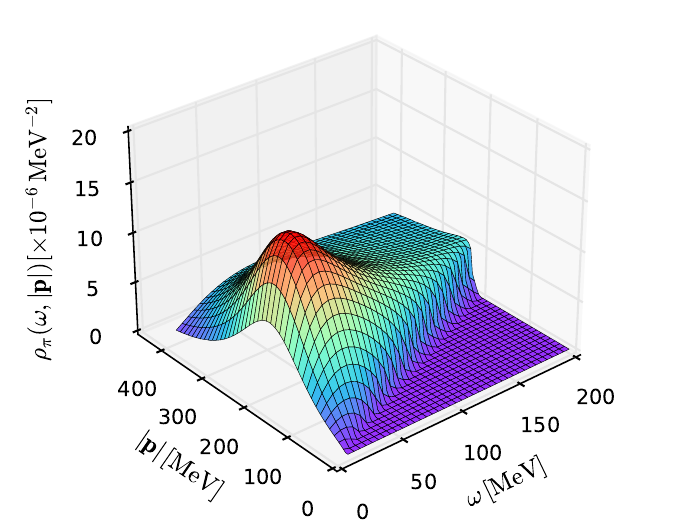}
\caption{3D plots of pion spectral function as a function of the frequency and spatial momentum at $\mu_B=0$, $T=160$ MeV (left panel) and $\mu_B=630$, $T=114$ MeV (right panel).}\label{fig:rhopi-3D}
\end{figure*}
%

%
\begin{figure*}[t]
\includegraphics[width=0.4\textwidth]{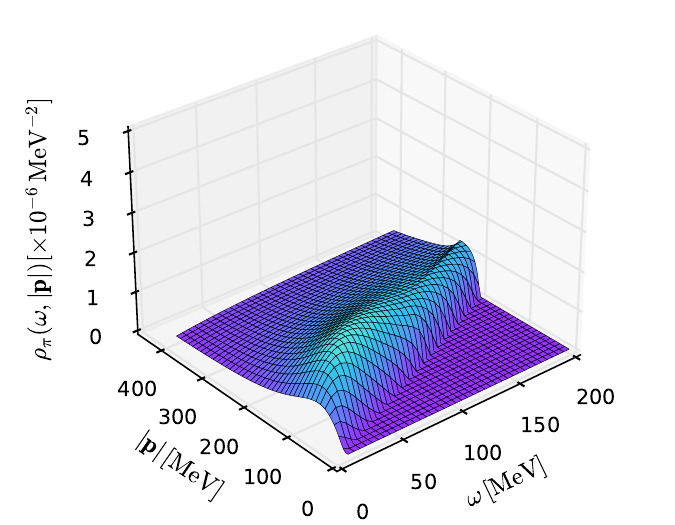}\hspace{0.5cm}
\includegraphics[width=0.4\textwidth]{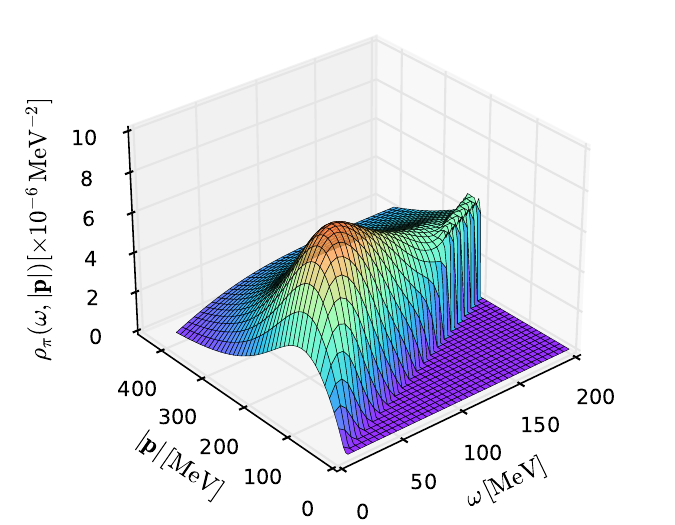}
\includegraphics[width=0.4\textwidth]{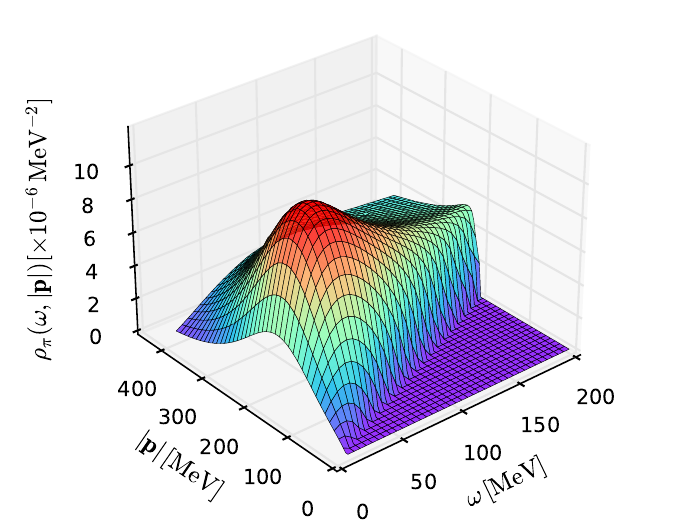}
\includegraphics[width=0.4\textwidth]{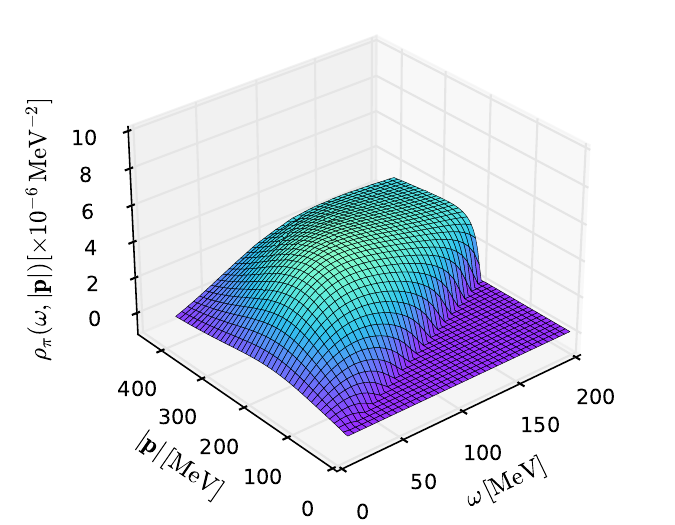}
\caption{3D plots of pion spectral function as a function of the frequency and spatial momentum at $\mu_B=550$ MeV, $T=100$ MeV (upper left panel), $\mu_B=550$ MeV, $T=120$ MeV (upper right panel), $\mu_B=550$ MeV, $T=130$ MeV (lower left panel) and $\mu_B=550$, $T=150$ MeV (lower right panel).}\label{fig:rhopi-3D-mub550}
\end{figure*}
%

In contrast, in the spacelike region there is a distinct difference between the normal and the moat regime. As also seen in the left plot of \Fig{fig:Gam2R}, in the moat regime there is a pronounced local minimum at nonzero $\bs{p}$ that extends towards finite $\omega$ until it eventually disappears for the reasons explained above. This behavior is therefore characteristic for the moat regime. This is in line with our findings in \Sec{sec:PH} and the discussion in \App{app:LandauDamping}, as PH processes, which we identified as the origin of the moat regime, only occur in the spacelike region. 

Note that the real and imaginary parts of the retarded two-point correlation function are related to each other through the dispersion relation, also known as the Hilbert transformation, i.e.,
\begin{align}
    {\rm Re}\,\Gamma_{\pi,R}^{(2)}(\omega,\bs{p}) = \mathcal{P}\int^{\infty}_{-\infty}\frac{d \omega'}{\pi}\frac{{\rm Im}\,\Gamma_{\pi,R}^{(2)}(\omega',\bs{p})}{\omega'-\omega}\,,
\end{align}
where $\mathcal{P}$ stands for the principal integral. From this dispersion relation, one can see that with the increase of $\omega$, even larger $|\bs{p}|$ is required to encode the PH fluctuations. This accounts for the results observed in the right panel of \Fig{fig:Gam2R-3D}, where the ridge deviates from the light cone gradually with the increase of $\omega$.

Now that we understood the retarded two-point function, we will discuss the pion spectral function. In \Fig{fig:rhopi-3D} we show $\rho_\pi(\omega,|\bs{p}|)$ in the normal regime at $(T,\mu_B) = (160,0)$\,MeV (left) and in the moat regime at $(T,\mu_B) = (114,630)$\,MeV (right), so for the same points in the phase diagram as in \Fig{fig:Gam2R-3D}. Since we established that the moat regime manifests itself in the spacelike region, we focus on frequencies $\omega\leq 200$\, MeV, where the relevant effect is most clearly seen. For these low frequencies, the spectral function is zero in the timelike region and the light cone is clearly visible through the onset of spacelike processes. The pion particle peaks from the zero crossings of ${\rm Re}\,\Gamma_{\pi,R}^{(2)}$ in \Fig{fig:Gam2R-3D} appears at larger $\omega$. 

We find a pronounced peak around $\omega \approx 50$\,MeV and $|\bs{p}| \approx 200$\,MeV in the moat regime which leads to a substantial enhancement of the spectral weight in this region. This enhancement is induced by the spacelike minimum of the real part of the retarded two-point function in the moat regime shown in \Fig{fig:Gam2R-3D}. This peak shows that in addition to the normal pion mode, there is another relevant contribution to the pion spectrum that may be attributed to a spacelike quasiparticle, the \emph{moaton}. In the following, we will confirm that this peak is indeed a feature of the moat regime, so calling it a moaton is appropriate.

%
\begin{figure}[t]
\includegraphics[width=0.45\textwidth]{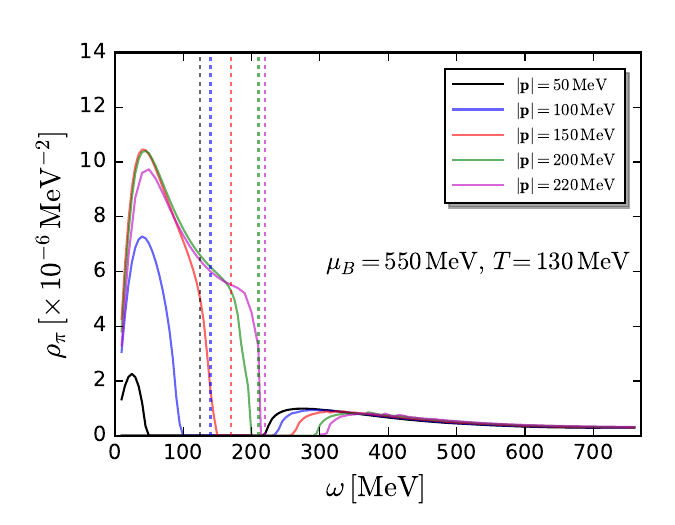}
\caption{Pion spectral function as function of the frequency for different spatial momenta $|\bs{p}|=$50, 100, 150, 200 and 220 MeV, at $\mu_B = 550$\,MeV and $T=130$\,MeV.}\label{fig:spec_mub550_T130_omega}
\end{figure}
%

\begin{figure}[t]
\includegraphics[width=0.4\textwidth]{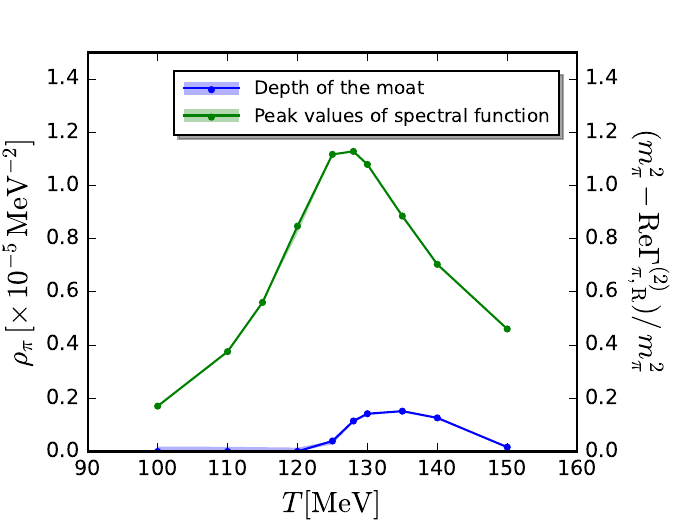}
\caption{Height of the moaton peak in the pion spectral function (green) and the depth of the moat (blue) as functions of $T$ at $\mu_B=550$ MeV. The depth of the moat is defined in \Eq{eq:dm}.}\label{fig:T-peak-mub550}
\end{figure}

In order to see how the moaton peak develops in the phase diagram, we show the pion spectral function at $\mu_B = 550$\,MeV and different $T$ in \Fig{fig:rhopi-3D-mub550}. The different temperatures in relation to the corresponding value of the spatial pion wave function are shown as the red dots in \Fig{fig:Zpion-T-mub550}. So in \Fig{fig:rhopi-3D-mub550} we go from outside the moat regime (top left plot), over just at the boarder of the moat regime (top right plot), and deep in the moat regime (bottom left plot), to even deeper into the moat regime (bottom right plot). We find that as the system enters the moat regime, a pronounced moaton peak quickly develops. Deeper into the moat regime, i.e.\ larger $T$ at fixed $\mu_B$, the overall magnitude of the spacelike contributions to the spectral function increases and, while there is still a peak, it only leads to a minor enhancement. This can be attributed to the fact that in the chirally restored phase, as $T$ increases, the pion mass rapidly increases and pions decouple from the QCD spectrum \cite{Fu:2019hdw}. Furthermore, at larger $T$ it is more likely to excite a quark/quark-hole from the heat bath, leading to an overall increase in the magnitude of spacelike processes. The height of the moaton peak is determined by the depth of the moat, i.e.\ the difference between static pion gap and the moaton gap, $E_\pi^{\rm (stat)}(\bs{0})-E_\pi^{\rm (stat)}(\bs{p}_M)$, cf.\ \Fig{fig:disper-relation}. The larger this difference, the higher the moaton peak. This will be discussed in more detail below; see also \Fig{fig:T-peak-mub550}. 

We note that in particular in the top right of \Fig{fig:rhopi-3D-mub550} one sees a spiked ridge structure close to the light cone at larger $\omega$. This arises because the timelike pion dispersion actually slightly crosses the light cone. In Ref.\ \cite{Yokota:2017uzu} this ``tachyonic mode" has been speculated to be related to an instability. But we emphasize that this is merely an artifact of the procedure used to extract the spectral function. As explained in \Sec{sec:FRG}, this procedure can be viewed as the first step of an iteration toward a self-consistent determination of the full momentum dependence of the meson propagator. This effect just shows that this iteration is not fully converged after just a single iteration step. We defer a self-consistent determination to future work but note that the main results here are unlikely to be affected by this. 

In order to get a complete picture, we show the full spectral function, including large frequencies, in \Fig{fig:spec_mub550_T130_omega} at $T=130$\,MeV and $\mu_B = 550$\,MeV. We clearly see three distinct contributions: the spacelike part from PH fluctuations, the pion particle pole (dashed line) from the zero-crossings of ${\rm Re}\,\Gamma_{\pi,R}^{(2)}$, and the timelike contributions dominated by the the decay of pions into quark-antiquark pairs. Most remarkably, the the moaton peak at nonzero spatial momentum leads to an enhancement by about an order of magnitude of the spacelike contribution over the timelike contribution. Since the spectral weight leads to experimental signatures from in-medium modifications \cite{Pisarski:2021qof, Rennecke:2023xhc, Nussinov:2024erh}, this enhancement could be the key to the experimental discovery of the moat regime.

%
\begin{figure*}[t]
\includegraphics[width=0.4\textwidth]{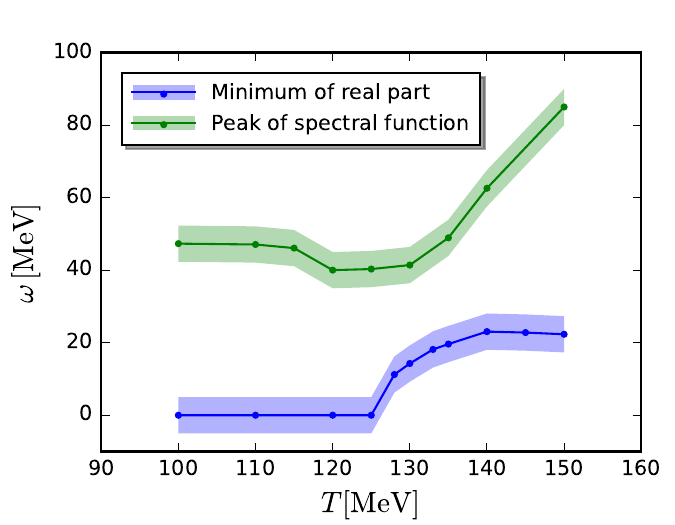}\hspace{0.5cm}
\includegraphics[width=0.4\textwidth]{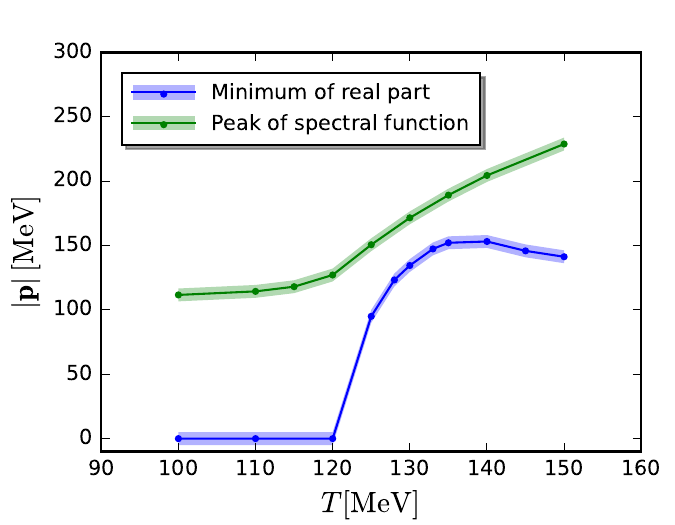}\hspace{0.5cm}
\caption{The behavior of $\omega$ (left) and $|\mathbf{p}|$ (right), corresponding to the minimum in the space-like region of real part of the two-point functions (blue), and the moaton peak of spectral functions (green), as functions of temperature at $\mu_B=550$ MeV.}\label{fig:omega-ps-peak-mub550}
\end{figure*}
%

%
%

While we have seen that the moat regime manifests itself in a clean and distinctive way in the real part of the retarded two-point function, the moaton peak is intertwined with the overall contribution from spacelike processes in the spectral function. To see the relation between the location of the spacelike minimum in ${\rm Re}\,\Gamma_{\pi,R}^{(2)}$ and the location of the moaton peak in the spectral function, we show both locations in \Fig{fig:omega-ps-peak-mub550} as functions $T$, again at $\mu_B = 550$\,MeV. The left plot shows the frequency at the minimum and the right plot shows the corresponding spatial momentum. The errors on these plots stem from our finite frequency and momentum resolution. We compute both $\omega$ and $|\bs{p}|$ in steps of 5 MeV, giving rise to the corresponding errors. 

As seen in \Fig{fig:Zpion-T-mub550}, the moat regime is entered at $T= 120$\,MeV. This is clearly reflected in the right plot of \Fig{fig:omega-ps-peak-mub550}, where the spatial momentum of the minimum of ${\rm Re}\,\Gamma_{\pi,R}^{(2)}$ becomes nonzero for $T\geq 120$\,MeV. $|\bs{p}|$ grows quickly until $T\approx 135$\,MeV, where it peaks at a value of about 150 MeV. Beyond this point, the minimum slowly approaches zero again with increasing $T$. This follows from asymptotic freedom. The location of the moaton peak in the spectral function follows this behavior, although more smoothly. Importantly, even outside the moat regime the spectral function has a maximum at nonzero $|\bs{p}|$ in the spacelike region. This is expected, as the spacelike processes build up continuously starting from the light cone and then eventually decrease with increasing momentum as it becomes unlikely to excite a quark/quark-hole with large momentum from the heat bath. In the moat regime, the moaton peak moves to larger spatial momenta on the order of 250 MeV, until it eventually disappears into the spacelike continuum when the pion mass becomes very large.   

In the left plot of \Fig{fig:omega-ps-peak-mub550} we show the frequency where the minimum of ${\rm Re}\,\Gamma_{\pi,R}^{(2)}$ and the moaton peak in $\rho_\pi$ are located. Similar to the spatial momentum in the right plot, and taking into account our error in the determination of $\omega$, the frequency of the minimum of ${\rm Re}\,\Gamma_{\pi,R}^{(2)}$ becomes nonzero in the moat regime. It peaks around $T \approx 140$\,MeV and then slowly returns to zero as $T$ increases further. The frequency of the moaton peak also grows in the moat regime. We conclude that there is a clear correlation between the moat regime and the position of the spacelike peak in the spectral function. The peak position is an important quantity, as its sets the momentum scale relevant for experimental searches of the moat regime.

The spectral functions in Figs.\ \ref{fig:rhopi-3D} and \ref{fig:rhopi-3D-mub550} show that the height of the peak in the spacelike region of the spectral function appears to be a distinctive feature of the moat regime. To corroborate this, we show the height of the moaton peak (the green line) as a function of $T$ at $\mu_B = 550$\, MeV in \Fig{fig:T-peak-mub550}. We can see that the peak reaches its maximum at $T\approx 130$\,MeV. The height of the peak is controlled by the depth of the moat, $d_M$, which can be defined as
\begin{align}\label{eq:dm}
   d_M = \frac{m_\pi^2-{\rm Re}\,\Gamma_{\pi,R}^{(2)}(\omega_{\rm min},\bs{p}_{\rm min})}{m_\pi^2}\,,
\end{align}
where $m_\pi^2 = {\rm Re}\,\Gamma_{\pi,R}^{(2)}(0,\bs{0})$ and $\omega_{\rm min}$ and $\bs{p}_{\rm min}$ are the frequency and spatial momentum at the minimum of the real part of the retarded two-point function. This is shown by the blue line in \Fig{fig:T-peak-mub550}. The depth characterizes to what extent finite-momentum excitations are favored over zero momentum excitations. It can hence be viewed as a measure of the relative importance of the moaton in spacelike pion spectrum. Since the depth also peaks around $T\approx 130$\,MeV, we find a clear connection to the height of the spacelike peak in the spectral function.

In summary, we found that the pion spectral function develops a new peak in the spacelike region in the moat regime. We have shown that it can be identified as a quasiparticle, the moaton, which controls the physics of the moat regime.

\subsection{Stability analysis}
\label{sec:stabilityana}

As discussed in the introduction and in \Sec{sec:spam}, since the moaton is spacelike and its mass gap is located at nonzero spatial momentum, a vanishing gap (i.e.\ a zero in the static dispersion relation) could indicate an instability towards the formation of a different, ordered phase. In the spectral function, such an instability would correspond to a very sharp moaton peak. This is equivalent to the situation at the CEP, where the massless critical mode gives rise to such a peak on the light cone \cite{Fujii:2003bz}, indicating the instability of the chirally symmetric phase towards a homogeneously broken phase. Since the minimum of the static dispersion is at nonzero $|\bs{p}| = |\bs{p}_M|$, the resulting instability would indicate that an inhomogeneous phase with a spatial modulation with wavenumber $\bs{p}_M$ is favored. 

To properly investigate the chiral phase structure, rather than the pions, the $\sigma$ meson is the relevant mode, as it carries the chiral condensate. By following the steps outlined in \Sec{sec:FRG}, we can obtain the fully momentum dependent $\sigma$ propagator in QCD exactly the same way as for the pions. For now, we are only interested in the phase structure and thus focus on the static properties. We will study the real time $\sigma$ correlations elsewhere.

Crucially, we find that the spatial wave function $Z_\sigma^\perp$ also turns negative at larger $\mu_B$. So the moat regime also affects scalar mesons. This is not much of a surprise, as the same PH processes that induce the moat behavior in pions happen for \emph{all} mesons. What is surprising, though, is that the moat regimes for $\sigma$ and pions are not identical. This can be seen in \Fig{fig:phase_moat}, where the chiral phase boundary of QCD, together with the CEP and moat regimes extracted from pion and $\sigma$ correlations, is shown. While the pion moat regime reaches slightly into the chirally broken phase and also covers the CEP, the $\sigma$ moat regime is shifted towards larger $T$ so that an overlap with the CEP is narrowly avoided. This confirms similar findings in the quark-meson model \cite{BPT}. This model is the emergent low energy effective theory the present functional QCD setup with emergent scalar-pseudoscalar mesons flows into; for a discussion see Ref.\ \cite{Dupuis:2020fhh, Fu:2022gou}. 

This observation itself may already indicate that there is no instability towards an inhomogeneous phase. At the CEP, there is an exactly massless critical mode. In our case, as we take into account nontrivial Polyakov loops \cite{Fu:2019hdw}, this mode is a mixture of the $\sigma$ mode and the Polyakov loops \cite{Haensch:2023sig}. Since it was demonstrated that the $\sigma$ is the dominant contribution to the critical mode, it becomes very light at the CEP. Thus, already a small $Z_\sigma^\perp < 0$ can induce an instability. In fact, a homogeneous CEP and moat behavior in the critical mode must be mutually exclusive, as otherwise any negative spatial wave function would always lead to an instability. This immediately follows from \Eq{eq:inst} for $m_\phi=0$. In Ref.\ \cite{Haensch:2023sig} it was shown that a homogeneous CEP and spatial modulations in the correlations of the critical mode induced by a so-called complex phase are mutually exclusive. While the relation between the moat regime and the complex phase is not fully understood yet, see also \cite{Schindler:2021otf, Nussinov:2024erh}, these phenomena are clearly related near the CEP. 

%
\begin{figure}[t]
\includegraphics[width=0.45\textwidth]{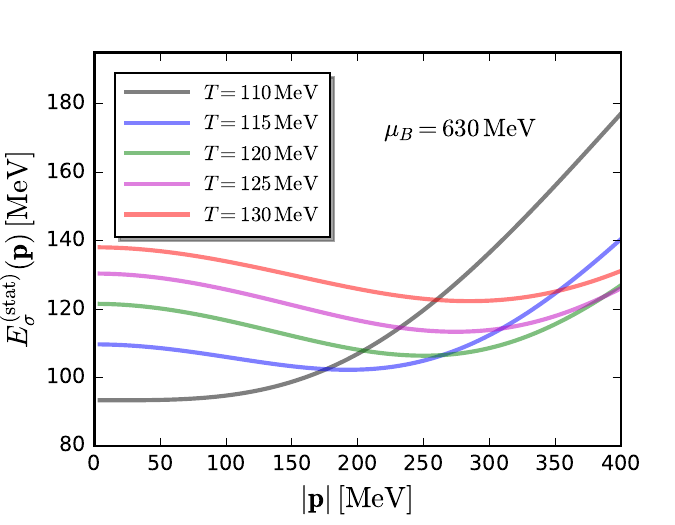}
\caption{Two-point correlation function of Sigma as functions of spatial momentum at vanishing external frequency. These curves are computed for a baryon chemical potential of 630 MeV at different temperatures.}\label{fig:disper-relation-sigma}
\end{figure}
%

In any case, we can also perform a direct stability analysis here. Since we did not find a vanishing moaton gap in the pion, we can already conclude that there is no instability in the pion mode here. This leaves us with the $\sigma$ mode. In \Fig{fig:disper-relation-sigma} we show the static sigma dispersion $E_\sigma^{\rm (stat)}$ as a function of spatial momentum for the largest available chemical potential in our study, $\mu_B = 630$\,MeV, and various temperatures. We note that this chemical potential is slightly below the critical $\mu_B$ of the CEP. At the lowest temperature in the figure, $T=110$\,MeV. the system is at the edge of the moat regime, and the $\sigma$ mass, $m_\sigma\approx 153$ MeV, is smaller than the pion mass at this point in the phase diagram. $E_\sigma^{\rm (stat)}$ is monotonically increasing with $|\bs{p}|$ in this case, but already very shallow at low momentum. At larger $T$ a minimum at nonzero $|\bs{p}|$ develops, so the system enters the $\sigma$ moat regime. With increasing $T$ the minimum becomes deeper and shifts to larger momenta. However, $m_\sigma$ also increases and the system is far away from an instability, since $E_\sigma^{\rm (stat)}(|\bs{p}|) > 0$ for all momenta and temperatures at $\mu_B = 630$\,MeV.

We have explicitly checked that this is also true at lower $\mu_B$, so we can exclude the occurrence of an inhomogeneous instability in the QCD for all $\mu_B \leq 630$\,MeV. As already mentioned in \Sec{sec:PH}, for a reliable stability analysis at larger $\mu_B$ and smaller $T$ we need to improve the systematics of our approach in this region. This is work in progress.


\section{Conclusion}

In order to understand the formation of spatially modulated phases in QCD, we have studied the underlying mechanism and the real time properties of the moat regime from first principles using the fRG. We have shown that the moat regime arises from particle-hole fluctuations of quarks at baryon chemical potentials $\mu_B \gtrsim 430$\,MeV above and around the pseudocritical temperature of the chiral transition. In fact, there always is a competition between particle-hole and creation-annihilation processes, where only the latter can lead to a moat regime. With increasing baryon density, the phase space in the vicinity of Fermi surface, where particle-hole fluctuations take place, increases as well. 
Consequently, these effects become more and more pronounced, and eventually dominate over creation-annihilation processes. It is hence natural to expect that the moat regime requires a sufficiently large $\mu_B$.

The real-time properties of the moat regime have been investigated in detail through the pion spectral function. Since particle-hole fluctuations are only kinematically allowed for spacelike mesons, they exclusive contribute to the spacelike region of the spectral function. In this region, we discovered a characteristic quasiparticle-like peak in the moat regime. We have demonstrated that this peak is a manifestation of the moat regime and hence dubbed this quasiparticle moaton. This retroactively justifies the use of this term in \cite{Nussinov:2024erh}.

We found that the moaton yields a substantial contribution to the spectral function, leading to distinct qualitative differences compared to the spectral function in absence of spatial modulations. This is encouraging for experimental searches for the moat regime and inhomogeneous phases in general, since the corresponding observables developed in Refs.\ \cite{Pisarski:2021qof, Rennecke:2023xhc, Nussinov:2024erh} all rely on the pion spectral function. Hence, characteristic effects in the spectral function will likely lead to characteristic signatures of the moat regime in observables such as Hanbury-Brown--Twiss correlations and dilepton production rates. We therefore plan to use our results to refine the predictions for future heavy-ion experiments, e.g., for CBM at FAIR.  

Furthermore, we also performed the first stability analysis directly in QCD. We have shown that instabilities towards the formation of an inhomogeneous phase can be deduced from the mass gap of moatons, i.e.\ the minimum of the static meson dispersion. Our results show that such instabilities are highly unlikely for $\mu_B \leq 630$\,MeV at any temperature. The method presented here, which relies on the dynamical hadronization technique, is complementary to the methods that have recently been developed in Refs.\ \cite{Motta:2023pks, Motta:2024agi, Motta:2024rvk}, which are geared towards studies of QCD using Dyson-Schwinger equations.

An open question related directly to the findings in Ref.\ \cite{Haensch:2023sig} is if and how in-medium mixing affects the moaton and potential inhomogeneous instabilities. It is conceivable that similar to the homogeneous instability at the CEP, the study of inhomogeneous instabilities requires a careful identification of the the relevant ``critical modes". We note that the results in the present work are not affected by this, since the system is so far away from an instability that it is unlikely that mixing will change any of our conclusions here. In general, the interplay between the critical mode of the CEP and the moaton is an interesting question that emerges from our work. One can imagine that near the CEP there are two peaks in the spacelike region of the spectral function of the critical mode, one from the moaton and the other from the critical mode. Since the $\sigma$ meson mode is the dominant contribution to the critical mode \cite{Haensch:2023sig}, we expect to see such features already in the $\sigma$ spectral function. 

Lastly, since the particle-hole fluctuations are not specific to QCD and play an important role in many condensed matter systems, 
our findings might also be of relevance there.

\section{Acknowledgements}
We thank the members of fQCD collaboration \cite{fQCD} for discussions and collaborations on related projects. We are grateful to L.\ von Smekal  and C.\ S.\ Fischer for numerous valuable discussions.  R.D.P.\
thanks T.\ Kunihiro for discussions.
S.Y.\ is supported by the Alexander v.\ Humboldt Foundation.
F.R.\ and S.Y.\ acknowledge support by the Deutsche Forschungsgemeinschaft (DFG, German Research Foundation) through the CRC-TR 211 ``Strong-interaction matter under extreme conditions"– project number 315477589 – TRR 211.
R.D.P.~is supported by the U.S.~Department
of Energy under contract DE-SC0012704, and thanks
the Alexander v.\ Humboldt Foundation for their support.
W.J.F.\ is supported by the National Natural Science Foundation of China under Grant Nos.\ 12175030, 12447102. 
J.M.P.~is funded by the Deutsche Forschungsgemeinschaft (DFG, German Research Foundation) by the Collaborative Research Centre SFB 1225 - 273811115 (ISOQUANT) and Germany’s Excellence Strategy EXC 2181/1 - 390900948 (the Heidelberg STRUCTURES Excellence Cluster). 
R.W.\ is supported by the Fundamental Research Funds for the Central Universities No.E3E46301.


\appendix



\section{Flow equations of mesonic two-point correlation functions and analytic continuation}
\label{app:flow}

\begin{figure}[t]
\includegraphics[width=1\columnwidth]{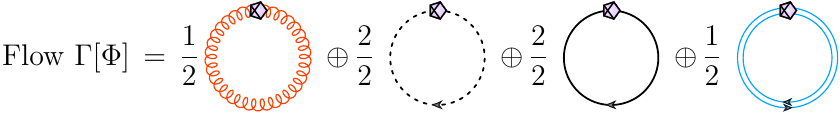}
\caption{Flow equation of the QCD effective action, where the loops denote the contributions from the gluon, ghost, quark, and meson fields, respectively. The lines stand for their full propagators and the crossed circles indicate infrared regulators.}\label{fig:actionflows-QCD}
\end{figure}

As outlined in \Sec{sec:FRG}, we use the fRG approach for QCD here. 
The diagrammatic representation of the flow of the QCD effective action with dynamical hadronization is shown in \Fig{fig:actionflows-QCD}. Flow receives contributions from the fluctuations of the gluon, ghost, quark, and meson fields in Landau gauge QCD, see \cite{Fu:2019hdw} for more details. 

\begin{figure*}[t]
\includegraphics[width=0.7\textwidth]{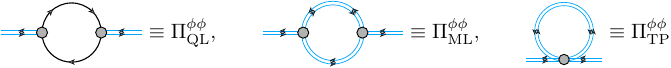}
\caption{Three different contributions to meson polarisations from the quark loop, meson loop, and the tadpole, respectively, see also \Fig{fig:flowsGam2phiphi}.}\label{fig:meson-polarisation}
\end{figure*}

The mesonic two-point function is obtained by a second derivative of the QCD effective action with respect to the emergent composite meson fields, 
\begin{align}
  \Gamma^{(2)}_{\phi \phi, k}=& \frac{\delta^2 \Gamma_k[\Phi]}{\delta \phi \delta \phi}\bigg|_{\Phi=\Phi_{\mathrm{EoM}}}\,,
  \label{eq:Gam2phiphi}
\end{align}
where $\Phi_{\mathrm{EoM}}$ denotes the fields on their respective equations of motion. Its flow is readily obtained from that of the effective action in \Fig{fig:actionflows-QCD}, upon implementing the relevant functional derivatives on both sides, and the resulting flow equation is depicted in \Fig{fig:flowsGam2phiphi}. Then the flow of mesonic two-point function at finite external momentum reads

\begin{align}
  &\partial_t\Gamma^{(2)}_{\phi\phi,k}(p)=\tilde{\partial}_t\bigg(-\big(\Pi^{\phi\phi}_{\mathrm{QL}}\big)_{ab}+\frac{1}{2}\big(\Pi^{\phi\phi}_{\mathrm{ML}}\big)_{ab}+\frac{1}{2}\big(\Pi^{\phi\phi}_{\mathrm{TP}}\big)_{ab}\bigg)\,.
  \label{eq:gammapi_Pi}
\end{align}
On the r.h.s.\ of the equation there are three different contributions resulting from the quark loop, meson loop, and the tadpole of mesons. These are denoted by $\Pi^{\phi\phi}_{\mathrm{QL}}$, $\Pi^{\phi\phi}_{\mathrm{ML}}$ and $\Pi^{\phi\phi}_{\mathrm{TP}}$, respectively, as shown in \Fig{fig:meson-polarisation}. 
The quark loop reads 
\begin{align}
  \big(\Pi^{\phi\phi}_{\mathrm{QL}}\big)_{ab}(p)&=\int \frac{d^4 q}{(2\pi)^4}\mathrm{Tr}\Big[h_k \tau^{a}G_{q}(q)h_k \tau^{b}G_{q}(q-p)\Big]\,,\label{eq:Piphi-QL}
\end{align}
where we use the subscripts $a, b$ to label the meson fields $\phi_{a}$ with $\phi_0=\sigma$ and  $\phi_i=\pi_i$ $(i=1,2,3)$. We use a $k$-dependent, but momentum-independent Yukawa coupling $h_k$. 
The matrices involved in the Yukawa interaction read $\tau^{0}=T^0$ and $\tau^{i}=i \gamma_5 T^i$, where $T^i$ $(i=1,2,3)$ are the generators of the group $SU(N_f=2)$ in flavor space with $\mathrm{Tr} (T^i T^j)=(1/2)\delta^{ij}$ and $T^{0}=(1/\sqrt{2N_{f}})\mathbbm{1}_{N_{f}\times N_{f}}$ with $N_f=2$. 
The quark propagator is given by
\begin{align}
  G_{q}(q)&=\frac{1}{Z_{q,k} i \Big[\gamma_0 q_0+\bs{\gamma}\cdot\bs{q}\big(1+r_q(\bs{q}^2/k^2)\big)\Big]+m_{q,k}} \,,\label{eq:Gq}
\end{align}
where a $3d$ flat regulator for the quark is used, i.e.,
\begin{align}
\begin{split}
  R_{\bar q q}&=Z_{q,k} i \bs{\gamma}\cdot\bs{q}\,r_q(\bs{q}^2/k^2)\,,\\[2ex] \mathrm{with} \quad r_q(x)&=\left(\frac{1}{\sqrt{x}}-1\right)\Theta(1-x)\,,\label{eq:Rq}
  \end{split}
\end{align}
for the convenience of calculations at finite temperature and densities. 
Here, $\Theta(x)$ stands for the Heaviside step function. 
Same as the Yukawa coupling, the quark wave function  $Z_{q,k}$ and the quark mass $m_{q,k}$ are dependent on the RG scale $k$. 
This $k$-dependence encodes most of the momentum dependence for masses and couplings, which in turn allows us to  neglect the subleading effect of the explicit momentum dependence, see \cite{Fu:2019hdw} for more discussions. 
Note that in the case of finite temperature and densities, the temporal component of the momentum in \labelcref{eq:Piphi-QL} is modified as $q_0\to q_0 +i \mu $, with the quark chemical potential $\mu$ related to the baryon chemical potential via $\mu=\mu_B/3$. $q_0$ is the Matsubara frequency, $q_0=(2n+1)\pi T $ for fermions here and $q_0=2n\pi  T$ for bosons with $n\in \mathbb{Z}$. 
Accordingly, the integral for $q_0$ in \labelcref{eq:Piphi-QL} is modified as a summation for the frequency.

The polarisations of $\sigma$ and $\pi$ mesons from the quark loop are readily obtained from \labelcref{eq:Piphi-QL} as well as \labelcref{eq:Gq}, which read 
\begin{align}
  &\big(\Pi^{\phi\phi}_{\mathrm{QL}}\big)_{00}(p)=-2 N_c \frac{h_k^2}{Z_{q,k}^2}\int \frac{d^4 q}{(2\pi)^4}\bar G_{q}(q)\bar G_{q}(q-p)\nonumber\\[2ex]
  &\times\Big[\Big(q_0(q_0-p_0)+\boldsymbol{q}\cdot(\boldsymbol{q}-\boldsymbol{p})\big(1+r_q(\boldsymbol{q}^2/k^2)\big)\nonumber\\[2ex]
  &\qquad\times\big(1+r_q((\boldsymbol{q}-\boldsymbol{p})^2/k^2)\big)-\bar m_{q,k}^2\Big)\Big]\,,\label{eq:Pisigma-QL} \\[2ex]
  &\big(\Pi^{\phi\phi}_{\mathrm{QL}}\big)_{ij}(p)=-2 N_c \frac{h_k^2}{Z_{q,k}^2}\delta_{ij}\int \frac{d^4 q}{(2\pi)^4}\bar G_{q}(q)\bar G_{q}(q-p)\nonumber\\[2ex]
  &\times\Big[\Big(q_0(q_0-p_0)+\boldsymbol{q}\cdot(\boldsymbol{q}-\boldsymbol{p})\big(1+r_q(\boldsymbol{q}^2/k^2)\big)\nonumber\\[2ex]
  &\qquad\times\big(1+r_q((\boldsymbol{q}-\boldsymbol{p})^2/k^2)\big)\Big)+\bar m_{q,k}^2\Big]\,,\label{eq:Pipi-QL} 
\end{align}
respectively. Here $N_c=3$ denotes the number of the colors. In \labelcref{eq:Pisigma-QL} and \labelcref{eq:Pipi-QL} one has $\bar m_{q,k}=m_{q,k}/Z_{q,k}$ and 
\begin{align}
  \bar G_{q}(q)&=\frac{1}{q_0^2+\boldsymbol{q}^2\big(1+r_q(\boldsymbol{q}^2/k^2)\big)^2+\bar m_{q,k}^2} \,.\label{eq:barGq}
\end{align}
%
%
%
%
It is also straightforward to evaluate the contribution of the meson loop to the mesonic two-point correlation functions, i.e., the second diagram in \Fig{fig:meson-polarisation}. 
One obtains for the $\sigma$ polarisation
\begin{align}
\nonumber
  &\big(\Pi^{\phi\phi}_{\mathrm{ML}}\big)_{00}(p)=\\[2ex] \nonumber
  &\quad\int \frac{d^4 q}{(2\pi)^4} G_{\sigma}(q) G_{\sigma}(q-p)\Big[ (2\rho)^3\big(V_k^{(3)}(\rho)\big)^2\\[2ex] \nonumber
  &\quad+6 (2\rho)^2 V_k^{(3)}(\rho) V_k^{(2)}(\rho)+9 (2\rho) \big(V_k^{(2)}(\rho)\big)^2\Big]\nonumber \\[2ex]
 &\quad+3\int \frac{d^4 q}{(2\pi)^4} G_{\pi}(q) G_{\pi}(q-p) (2\rho) \big(V_k^{(2)}(\rho)\big)^2\,,\label{eq:Pisigma-ML}
\end{align}
and for the $\pi$ polarisation
\begin{align}
  \big(\Pi^{\phi\phi}_{\mathrm{ML}}\big)_{ij}(p)=&\delta_{ij} 2\int \frac{d^4 q}{(2\pi)^4} G_{\sigma}(q) G_{\pi}(q-p) (2\rho) \big(V_k^{(2)}(\rho)\big)^2\,.\label{eq:Pipi-ML}
\end{align}
Note that the meson interactions, i.e., the three-meson vertices in \labelcref{eq:Pisigma-ML} and \labelcref{eq:Pipi-ML} as well as the four-meson vertex in the tadpole in \Fig{fig:meson-polarisation}, are described by a $k$-dependent mesonic effective potential $V_k(\rho)$ and its $n$-order derivatives $V_k^{(n)}(\rho)$. 
Here $V_k(\rho)$ is $O(4)$-invariant with $\rho=\phi^2/2$ for $N_f=2$ flavor light quarks. Evidently, in the setup of the present work the meson vertices on the r.h.s.\ of the flow equation in \Fig{fig:flowsGam2phiphi}, i.e., the three-meson vertices in the meson loop and the four-meson vertex in the tadpole, are momentum-independent but rather $k$-dependent, which share the same qualification as, e.g., the Yukawa coupling discussed above. 
More importantly, this truncation for couplings facilitates computations significantly, and makes the analytic continuation from the Euclidean to Minkowski regime in \labelcref{eq:AnalyContiGamphi2} in the main text possible on the level of the analytic flow equation. 
The meson propagators in \labelcref{eq:Pisigma-ML} and \labelcref{eq:Pipi-ML} read
\begin{align}
  G_{\phi}(q)&=\frac{1}{Z_{\phi,k} \Big[q_0^2+\boldsymbol{q}^2\big(1+r_{\phi}(\boldsymbol{q}^2/k^2)\big)\Big]+m_{\phi,k}^2} \,,\label{eq:Gphi}
\end{align}
with the $3d$ flat regulator for mesons given by
\begin{align}
  R_{\phi}&=Z_{\phi,k} \boldsymbol{q}^2\,r_{\phi}(\boldsymbol{q}^2/k^2)\,,\quad \quad r_{\phi}(x)=\left(\frac{1}{x}-1\right)\Theta(1-x)\,,\label{eq:Rphi}
\end{align}
where the curvature masses of mesons are related to the effective potential via the equations as follows
\begin{align}
  m_{\sigma,k}^2&=V_k^{\prime}(\rho)+2\rho V_k^{(2)}(\rho)\,,\quad \quad m_{\pi,k}^2=V_k^{\prime}(\rho)\,.
\end{align}
Finally, we present the contribution of the tadpole diagram to the meson polarisations, which are given by
\begin{align}
\nonumber
  \big(\Pi^{\phi\phi}_{\mathrm{TP}}\big)_{00}(p)=&-\int \frac{d^4 q}{(2\pi)^4} \Big[ G_{\sigma}(q)\Big( (2\rho)^2V_k^{(4)}(\rho)\\
  \nonumber
  &+6 (2\rho) V_k^{(3)}(\rho)+3 V_k^{(2)}(\rho)\Big)\\
  &+3 G_{\pi}(q)\Big( (2\rho) V_k^{(3)}(\rho)+V_k^{(2)}(\rho)\Big)\Big]\,,\label{eq:Pisigma-TP}\\[2ex]
  \nonumber
  \big(\Pi^{\phi\phi}_{\mathrm{TP}}\big)_{ij}(p)=&-\delta_{ij} \int \frac{d^4 q}{(2\pi)^4}\Big[ G_{\sigma}(q)\Big( (2\rho) V_k^{(3)}(\rho)\\
  &+V_k^{(2)}(\rho)\Big)+5 G_{\pi}(q)V_k^{(2)}(\rho)\Big]\,.\label{eq:Pipi-TP}
\end{align}
Obviously, the tadpole is independent of the external momentum $p$, since the four-meson vertex is momentum-independent as discussed above.

In summary, substituting \labelcref{eq:Pisigma-QL}, \labelcref{eq:Pipi-QL}, \labelcref{eq:Pisigma-ML}, \labelcref{eq:Pipi-ML}, \labelcref{eq:Pisigma-TP}, and \labelcref{eq:Pipi-TP} into the r.h.s. of flow equation in \Fig{fig:flowsGam2phiphi}, one is able to obtain the two-point Euclidean correlation functions for mesons. 
This flow equation also allows us to perform the analytic continuation, such that real-time functions can be achieved on the level of the flow equation. 
To that end, one also has to provide $k$-dependent quantities required on the r.h.s. of the flow equation, such as the Yukawa coupling $h_k$, the quark mass $m_{q,k}$, the quark wave function $Z_{q,k}$ or equivalently the quark anomalous dimension $\eta_{q,k}=-\partial_t Z_{q,k}/Z_{q,k}$, the meson wave function $Z_{\phi,k}$ or the meson anomalous dimension $\eta_{\phi,k}=-\partial_t Z_{\phi,k}/Z_{\phi,k}$, the mesonic effective potential $V_k(\rho)$, in this work all of which are imported from the first-principles calculations to QCD at finite temperature and densities within the fRG approach in \cite{Fu:2019hdw}, more details can be found there.

\section{Particle-hole fluctuations of quarks at large baryon chemical potentials}
\label{app:LandauDamping}

Since in this work we only calculate the spectral function of pion, we here present the flow equation for the pion two-point correlation function at finite external momentum. As the moat behavior of the two-point function occurs only in the symmetric phase at high chemical potential, where the contribution from quark loops dominates over that of meson loops, we can neglect the meson loop contribution here and present only the flow equation for the quark self-energy,
\begin{widetext}
\begin{align}
\nonumber
  &\tilde{\partial}_t\big(\Pi^{\phi\phi}_{\mathrm{QL}}\big)_{ij}(p)=-2N_c\frac{h^2_k}{k\,Z^2_{q,k}}\delta_{ij}\int\frac{d^3\boldsymbol{q}}{(2\pi)^3}\Big[1+(|\boldsymbol{q}|/k-1)\eta_{q,k}\Big]\Theta(1-\boldsymbol{q}^2/k^2)\nonumber\\[2ex]
  &\times\bigg\{-2\mathcal{F}_{(2)}+\Big[1+\frac{(\boldsymbol{q}-\boldsymbol{p})^2}{k^2}(1+r_{q}((\boldsymbol{q}-\boldsymbol{p})^2/k^2))^2+\frac{p^2_0}{k^2}-2\frac{\boldsymbol{q}\cdot(\boldsymbol{q}-\boldsymbol{p})}{k^2}\big(1+r_q(\boldsymbol{q}^2/k^2)\big)\big(1+r_q((\boldsymbol{q}-\boldsymbol{p})^2/k^2))\big)\Big]\mathcal{FF}^-_{(2,1)}\nonumber\\[2ex]
  &\qquad\qquad\quad\,\,\,+\Big[1+\frac{(\boldsymbol{q}+\boldsymbol{p})^2}{k^2}(1+r_q((\boldsymbol{q}+\boldsymbol{p})^2/k^2))^2+\frac{p^2_0}{k^2}-2\frac{\boldsymbol{q}\cdot(\boldsymbol{q}+\boldsymbol{p})}{k^2}\big(1+r_q(\boldsymbol{q}^2/k^2)\big)\big(1+r_q((\boldsymbol{q}+\boldsymbol{p})^2/k^2))\big)\Big]\mathcal{FF}^+_{(2,1)}\nonumber\\[2ex]
  &\qquad\qquad\quad\,\,\,+\Big[-1+\frac{\boldsymbol{q}\cdot(\boldsymbol{q}-\boldsymbol{p})}{k^2}\big(1+r_q(\boldsymbol{q}^2/k^2)\big)\big(1+r_q((\boldsymbol{q}-\boldsymbol{p})^2/k^2))\big)\Big]\mathcal{FF}^-_{(1,1)}\nonumber\\[2ex]
  &\qquad\qquad\quad\,\,\,+\Big[-1+\frac{\boldsymbol{q}\cdot(\boldsymbol{q}+\boldsymbol{p})}{k^2}\big(1+r_q(\boldsymbol{q}^2/k^2)\big)\big(1+r_q((\boldsymbol{q}+\boldsymbol{p})^2/k^2))\big)\Big]\mathcal{FF}^+_{(1,1)}\bigg\}\,.
  \label{eq:gammapi_FF}
\end{align}
\end{widetext}
In the flow equation we use the dimensionless fermion threshold functions, which represent the quark propagators summed over Matsubara frequencies. The function at vanishing momentum reads
\begin{align}
  \mathcal{F}_{(n)}\equiv k^{2n-1}\,T\sum_{n_q}\Big(\bar G_q(q, \bar m^2_{q})\Big)^n \,,\label{eq:Fn}
\end{align}
and the finite momentum one reads
\begin{align}
\nonumber
  &\mathcal{FF}^{\pm}_{(m,n)}(p)\equiv\\ 
  &\quad k^{2(m+n)-1}\,T\sum_{n_q}\Big(\bar G_q(q, \bar m^2_{q})\Big)^m \Big(\bar G_q(q\pm p, \bar m^2_{q})\Big)^n\,,\label{eq:FFmn}
\end{align}
with integers $m$, $n$ and
\begin{align}
  \bar G_{q}(q)&=\frac{1}{(q_0+i \mu)^2+\big(E(|\bs{q}|)\big)^2} \,,\label{eq: barGq2}
\end{align}
where one has
\begin{align}
  E(|\bs{q}|)&=\sqrt{\bs{q}^2\big(1+r_q(\bs{q}^2/k^2)\big)^2+\bar m_{q,k}^2}\,,
\end{align}
$q_0=(2n_q+1)\pi T $ and $\mu$ is the quark chemical potential. Since the fermion regulator can be inserted into either of the two different internal propagators separately in the quark loop, we consider two different momentum configurations $\bs{q}+\bs{p}$ and $\bs{q}-\bs{p}$ in the flow equation. Thus the corresponding threshold functions are labeled with $\pm$ signs. Here we focus on the simplest case with $m=n=1$ as this already captures the effects of interest here.

Performing the sum in \labelcref{eq:FFmn} yields two distinct contributions,
\begin{align}
  \mathcal{FF}^\pm_{(1,1)}(p)&= \mathcal{FF}_{(1,1)}^{\pm,\mathrm{CA}}(p)+\mathcal{FF}_{(1,1)}^{\pm,\mathrm{PH}}(p)\,,
\end{align}
with
\begin{widetext}
\begin{align}\label{eq:FFCA}
   \mathcal{FF}_{(1,1)}^{\pm,\mathrm{CA}}(p)=\frac{k^3}{4E(|\bs{q}|)E(|\bs{q}\pm\bs{p}|)}&\Bigg\{\frac{1}{i p_0-E(|\bs{q}|)-E(|\bs{q}\pm\bs{p}|)}\bigg[-1+n_F\big(E(|\bs{q}|);T,\pm\mu\big)+n_F\big(E(|\bs{q}\pm\bs{p}|);T,\mp\mu\big)\bigg]\nonumber\\[2ex]
    &+\frac{1}{i p_0+E(|\bs{q}|)+E(|\bs{q}\pm\bs{p}|)}\bigg[1-n_F(E(|\bs{q}|);T,\mp\mu)-n_F(E(|\bs{q}\pm\bs{p}|);T,\pm\mu)\bigg]\Bigg\}\,,
\end{align}
\begin{align}\label{eq:FFPH}
    \mathcal{FF}_{(1,1)}^{\pm,\mathrm{PH}}(p)=\frac{k^3}{4E(|\bs{q}|)E(|\bs{q}\pm\bs{p}|)}&\Bigg\{\frac{1}{i p_0-E(|\bs{q}|)+E(|\bs{q}\pm\bs{p}|)}\bigg[-n_F\big(E(|\bs{q}|);T,\mp\mu \big)+n_F\big(E(|\bs{q}\pm\bs{p}|);T,\mp\mu\big)\bigg]\nonumber\\[2ex]
    &+\frac{1}{i p_0+E(|\bs{q}|)-E(|\bs{q}\pm\bs{p}|)}\bigg[n_F\big(E(|\bs{q}|);T,\pm\mu\big)-n_F\big(E(|\bs{q}\pm\bs{p}|);T,\pm\mu\big)\bigg]\Bigg\}\,,
\end{align}
\end{widetext}
where the fermionic distribution function reads
\begin{align}
  n_F(E;T,\mu)&=\frac{1}{\exp\Big[(E-\mu)/T\Big]+1}\,.
\end{align} 
We take into account the Polyakov loops in this work, which lead to modifications of these distribution functions; see, e.g., Eq.\ (N8) in \cite{Fu:2019hdw}. 

The first contribution, $\mathcal{FF}^{\mathrm{CA}}$, is a genuinely relativistic contribution that describes the creation and annihilation of quarks and antiquarks. 
$\mathcal{FF}^{\mathrm{PH}}$ stems from particle-hole fluctuations of quarks which also exist in the nonrelativistic limit of the theory.
This can be seen as follows: The scalar part of the quark propagator $\bar G_q$ in \Eq{eq: barGq2} can be written as
\begin{align}\label{eq:barGPA}
    \bar G_q(q) = \frac{1}{(iq_0)-\big[ E(|\bs{q}|-\mu) \big]}\cdot\frac{1}{(iq_0)-\big[ -E(|\bs{q}|-\mu) \big]}
\end{align}
The first factor is the particle contribution which corresponds to the upper, positive-energy Dirac cone. The second factor is the antiparticle contribution that corresponds to the lower, negative energy Dirac cone. This is illustrated in \Fig{fs}. As usual, the sum in \Eq{eq:FFmn} is expressed as a sum over the residues at these particle and antiparticle poles. This leads to Eqs.\ \labelcref{eq:FFCA} and \labelcref{eq:FFPH}. 
It is straightforward to verify that all terms in these equations with factors that contain energy sums,
\begin{align}
    \frac{1}{p_0\pm\big[ E(|\bs{q}|) +E(|\bs{q}\pm\bs{p}|) \big]}\,,
\end{align}
arise from the residues involving one particle and one antiparticle. Owing to the Pauli principle, this process stems from the creation/annihilation (CA) processes of antiparticles in the Dirac sea of the negative-energy Dirac cone and particles above the Fermi surface. The terms containing energy differences, 
\begin{align}
    \frac{1}{p_0\pm\big[ E(|\bs{q}|) -E(|\bs{q}\pm\bs{p}|) \big]}\,,
\end{align}
stem from particle-particle contributions. They arise from a ``hole" in the Dirac sea in the positive-energy Dirac cone and a particle above the Fermi surface and are hence called particle-hole (PH) fluctuations. In the nonrelativistic limit, there are no antiparticles, so the second factor in \Eq{eq:barGPA} is dropped and the PH-contribution is all that is left. 

The same decomposition can also be applied to higher-order functions $\mathcal{FF}_{(m,n)}$ with $(m,n) > (1,1)$, including the meson polarization in \labelcref{eq:Piphi-QL}. 

An equivalent way to understand these different contributions is to inspect the thermal distributions occurring in Eqs.\ \eq{eq:FFCA} and \eq{eq:FFPH}. The process where an off-shell meson $\phi^*$ creates a quark-antiquark pair, $\phi^*\rightarrow q + \bar q$, comes with a thermal factor $\big[1-n_F(E,\mu)\big]\big[1-n_F(\bar E,-\mu)\big]$,
because a quark with energy $E$ and an antiquark with energy $\bar  E$ can only be created if these states are unoccupied. The inverse annihilation process $ q + \bar q \rightarrow \phi$ involves occupied quark and antiquark states and hence comes with a factor $n_F(E,\mu) n_F(\bar E,-\mu)$. The total CA process must hence come with a factor
\begin{align}
\begin{split}
    &\big[1-n_F(E,\mu)\big]\big[1-n_F(\bar E,-\mu)\big]- n_F(E,\mu) n_F(\bar E,-\mu)\\
    &\quad= 1 -n_F(E,\mu) -n_F(\bar E,-\mu)\,.  
\end{split}
\end{align}
This is clearly the case for $\mathcal{FF}^{\mathrm{CA}}$ in \Eq{eq:FFCA}.

The other possible processes are the absorption, $\phi^* + q \rightarrow q$, and emission, $q \rightarrow q + \phi^* $, of an off-shell meson by a quark (and the same with antiquarks). One may also view this as a meson creating or annihilating a quark--quark-hole pair. For such a process to happen, one always needs an occupied and an unoccupied quark state. If $E_{1,2}$ are the energies of these states, these contributions get a thermal prefactor
\begin{align}
\begin{split}
      &n_F(E_1,\mu)\big[1-n_F(E_2,\mu)\big] -n_F(E_2,\mu)\big[1-n_F(E_1,\mu)\big]\\ 
      &\quad = n_F(E_1,\mu) - n_F(E_2,\mu)\,,
\end{split}
\end{align}
and the same for antiquarks. So the PH contribution $\mathcal{FF}^{\mathrm{PH}}$ in \Eq{eq:FFPH} describes these processes. They can only occur in a medium and are therefore also sometimes called \emph{Landau damping}.

Since the quarks are on-shell, PH processes are kinematically only possible for \emph{spacelike} mesons, i.e.\ if the meson momentum is larger than its frequency, $|\bs{p}|\geq \omega$. In contrast, CA processes are only possible for timelike mesons, as they need to create/annihilate on-shell quarks. For a comprehensive breakdown of the different thresholds of scattering processes, we refer, e.g., to Ref.~\cite{Jung:2016yxl}.

\bibliography{ref-lib}
    
\end{document}